\documentclass[apj, twocolumn]{emulatearxiv}
\usepackage{amsmath}
\usepackage{listings}

\begin{document}

\def\lsim{\mathrel{\rlap{\lower 4pt \hbox{\hskip 1pt $\sim$}}\raise 1pt
\hbox {$<$}}} 
\def\gsim{\mathrel{\rlap{\lower 4pt \hbox{\hskip 1pt $\sim$}}\raise 1pt
\hbox {$>$}}}

\title{Pulsational Pair-instability Supernovae. I. Pre-collapse
Evolution and Pulsational Mass Ejection}

\author{
Shing-Chi Leung\thanks{Email address: shingchi.leung@ipmu.jp},
}

\affiliation{Kavli Institute for the Physics and 
Mathematics of the Universe (WPI), The University 
of Tokyo Institutes for Advanced Study, The 
University of Tokyo, Kashiwa, Chiba 277-8583, Japan}

\affiliation{TAPIR, Walter Burke Institute for Theoretical Physics, 
Mailcode 350-17, Caltech, Pasadena, CA 91125, USA}

\author{Ken'ichi Nomoto\thanks{Email address: nomoto@astron.s.u-tokyo.ac.jp}}

\affiliation{Kavli Institute for the Physics and 
Mathematics of the Universe (WPI), The University 
of Tokyo Institutes for Advanced Study, The 
University of Tokyo, Kashiwa, Chiba 277-8583, Japan}

\author{Sergei Blinnikov\thanks{Email address: sblinnikov@gmail.com}}

\affiliation{Kavli Institute for the Physics and
Mathematics of the Universe (WPI), The University
of Tokyo Institutes for Advanced Study, The
University of Tokyo, Kashiwa, Chiba 277-8583, Japan}

\affiliation{NRC ``Kurchatov Institute'' -- ITEP, B.Cheremushkinkaya 25, 117218 Moscow, Russia}

\affiliation{Dukhov Automatics Research Institute (VNIIA), Suschevskaya 22, 127055 Moscow, Russia}
\date{\today}

-------------------------------------------------------------------


\begin{abstract}

We calculate the evolution of massive stars, which undergo pulsational
pair-instability (PPI) when the O-rich core is formed.  The evolution
from the main-sequence through the onset of PPI is calculated for
stars with the initial masses of 80 -- 140 $M_{\odot}$ and
metallicities of $Z = 10^{-3} - 1.0$ $Z_\odot$.  Because of mass loss,
$Z \leq 0.5$ $Z_\odot$ is necessary for stars 
to form  He cores massive enough (i.e., mass $>40 ~M_\odot$) to undergo PPI.
The hydrodynamical phase of evolution from PPI through
the beginning of Fe core collapse is calculated for the He cores with
masses of $40 - 62 ~M_\odot$ and $Z = 0$.  During PPI,
electron-positron pair production causes a rapid contraction of the
O-rich core which triggers explosive O-burning and a pulsation of the
core.  We study the mass dependence of the pulsation dynamics,
thermodynamics, and nucleosynthesis.  The pulsations are stronger for
more massive He cores and result in such a large amount of mass
ejection such as 3 -- 13 $M_\odot$ for $40 - 62 ~M_\odot$ He cores.  These He
cores eventually undergo Fe-core collapse.  The $64 ~M_\odot$ He core
undergoes complete disruption and becomes a pair-instability supernova.  
The H-free circumstellar matter ejected
around these He cores is massive enough for to explain the observed
light curve of Type I (H-free) superluminous supernovae with
circumstellar interaction.  We also note that the mass ejection sets
the maximum mass of black holes (BHs) to be $\sim 50$ $M_{\odot}$,
which is consistent with the masses of BHs recently detected by VIRGO
and aLIGO.


\end{abstract}

\pacs{
26.30.-k,    
}

\keywords{stars: oscillations (including pulsations) -- (stars:) supernovae: general -- stars: evolution -- (stars:) circumstellar matter -- stars: black holes}


\section{Introduction}

\subsection{Pulsational Pair-Instability (PPI)}

The structure and evolution of massive stars depend on stellar mass,
metallicity, and rotation \cite[e.g.,][]{Arnett1996, Nomoto1988,
Heger2000, Heger2002, Nomoto2013, Meynet2017, Limongi2017,
Hirschi2017}.  In stars with the zero-age main-sequence (ZAMS) mass
of $M \sim 10-80 ~M_{\odot}$, hydrostatic burning progresses from
light elements to heavy elements in the sequence of H, He, C, O, Ne,
and Si burning, and finally a Fe core forms and gravitationally
collapses to form a compact object such as a neutron star.

For very massive stars with $M \gsim 80 ~M_\odot$ (i.e. a He core of mass
greater than 35 $M_{\odot}$), but exact correspondence is strongly 
metallicity dependent \citep{ElEid1983,Heger2002,Hirschi2017,Woosley2017}, effects of the
electron-positron pair-production ($\gamma \rightarrow e^- + e^+$) on
stellar structure and evolution are important when the O-rich core is
formed \citep{Fowler1964,Fraley1968}.  Pair-production causes the dynamically
unstable contraction of the O-rich core, which ignites explosive
O-burning.  For $140 ~M_\odot \lsim M \lsim 300 ~M_{\odot}$, 
the released nuclear energy is large enough to disrupt the whole star, so
that the star explodes as pair-instability supernovae (PISN)
\citep{Barkat1967,Bond1984,Baraffe2001}. Above $\sim 300 ~M_{\odot}$,
the star collapses to form black hole again. As a result, no black hole can be formed 
with a mass between $\sim 50 ~M_{\odot}$ and $\sim 150 ~M_{\odot}$ \citep{Heger2002}. 
Such mass gap may provide distinctive features in the mass spectrum of black hole
through the detection of merger event of binary black holes.

For stars with $M = 80 - 140 ~M_{\odot}$ [He cores of 
35 -- $\sim$ 65 $M_{\odot}$ \citep{Woosley2017}], explosive O burning does not
disrupt the whole star, but creates strong pulsations
\citep{Barkat1967, Rakavy1967}, which is called Pulsational 
Pair-instability (PPI).  These stars undergo distinctive
evolution compared to more massive or less massive stars.  PPI is
strong enough to induce massive mass ejection as in PISN, while the
star further evolves to form an Fe core that collapses into a compact
object later as a core collapse supernova (CCSN).  PPI supernovae
(PPISNe) of $80 - 140 ~M_{\odot}$ stars are thus the hybrid of PISN
and CCSN.  

The exact ZAMS mass range of PPISN depends on the mass loss by stellar
wind, thus on metallicity, and also on rotation.  For PPISN
progenitors, the wind mass loss during H and He burning phases could
contribute to the loss of almost a half of the initial progenitor
mass (See e.g. Table 2 of \cite{Woosley2017}). 
Such mass loss can suppress the formation of a massive He-core.
The exact mass of the He core as a function of metallicity and
ZAMS mass remains less understood because the mass loss
processes in massive star are not well constrained \citep{Renzo2017}.  
Rotation provides additional support by the centripetal force, which
allows PPISN to be formed at an even higher progenitor mass
\citep{Glatzel1985, Chatzopoulos2012}.

In order to pin down the mass range of PPISN, a mass survey of
main-sequence star models is done in \cite{Heger2002, Ohkubo2009} with
focus on the zero metallicity stars.  Large surveys in other
metallicity can be also found in e.g. \cite{Heger2010,
Sukhbold2016}.  A large array of stellar models covering also PPISN
with rotation has been further explored in \cite{Yoon2012}.

The evolution of PPISN is very dynamical in the late phase.  During
the pulsation, the dynamical timescale can be comparable with the
nuclear timescale that hydrostatic approximation is no longer a good
approximation. Also, when the star drastically expands after the
energetic nuclear burning triggered at the contraction, the subsequent
shock breakout near the surface is obviously a dynamical
phenomenon. This suggests that during this dynamical but short phase,
hydrodynamics instead of hydrostatic is required in order to follow
the evolution consistently.  Compared to hydrodynamical studies of
PISNe \citep{Barkat1967,Umeda2002,Heger2002,Scannapieco2005,Chatzopoulos2013,Chen2014},
systematic hydrodynamical study of PPI has been conducted only
recently \cite[e.g.,][]{Woosley2015,Woosley2017}.

\subsection{Connections to Observations}

The optical aspect of PPI and PPISN might explain some super-luminous
supernovae (SLSNe), such as SN2006gy \citep{Woosley2007, Kasen2011,
Chen2014}.  Recent modelling of SLSN PTF12dam \citep{Tolstov2017}
has required an explosion of a 40 $M_{\odot}$ star with 20 -- 40
$M_{\odot}$ circumstellar medium (CSM) with a sum of 6 $M_{\odot}$
$^{56}$Ni in the explosion. The shape, rising time and fall rate of
the light curves provide constraints on the composition, density and
velocity of the ejecta, which give insights to the modeling of
PPISN.  It demonstrates the importance to track the mass loss history
of a star prior to its collapse.  The rich mass ejection can be an
explanation to the dense CSM observed in some supernovae, such as SN
2006jc \citep{Foley2007}.  Supernova models in the PPISN mass range
are further applied to explain some unusual objects, including SN
2007bi \citep{Moriya2010, Yoshida2014} and iPTF14hls
\citep{Woosley2018}.

There is also a possible connection to the well observed Eta Carinae,
which has demonstrated significant mass loss of about 30 $M_{\odot}$
\citep{Smith2007,Smith2008}.

Furthermore, recent detections of the gravitational waves emitted by
the merging of black holes (BH) \citep{Abbott2016a, Abbott2016b}, such
as GW150914 and GW170729 imply existence of BHs of masses $\sim 30 - 50 ~M_{\odot}$.  In order to study the mass spectrum of BHs in this
mass range, the evolutionary path of this class of objects becomes
necessary.  Such observations have led to the interest in the
evolutionary origin of massive BHs, including PPI phenomena
\cite[e.g.,][]{Woosley2017, Belczynski2017, Marchant2018}.
Our calculations will update the lower end of the "mass gap" 
of the massive BHs (not near the NS-BH boundary).

\subsection{Present Study}

From the above importance of PPI, we re-examine PPI by using the
open-source stellar evolution code MESA (v8118; \cite{Paxton2011, Paxton2013, Paxton2015, Paxton2017}).  

We use this version because
the recent update of the code \citep{Paxton2015} has included an
implicit energy-conserving \citep{Grott2005} hydrodynamical scheme as
one of its evolution options.

We study a series of the evolution of stars from ZAMS for the masses
ranging from 80 to 140 $M_{\odot}$ and various metallicities. This
corresponds to the He-core masses from $\sim 40$ to $65 ~M_{\odot}$.
Then we calculate the evolution of such He stars to study the
hydrodynamical behaviour of PPI including mass ejection.

In Section \ref{sec:methods}
we describe the code for 
preparing the initial models and the details 
of the one-dimensional implicit hydrodynamics
code for the pulsation phase.

In Section
\ref{sec:MainSeq} we examine the evolutionary
path of PPISN in the H and He-burning phases
and the influence of metallicity on the final
He and CO-core masses. 

Then in Section \ref{sec:hydro},
we first present the pre-pulsation evolution of our models which
includes
He- and C-burning phases. We study the dynamics
of the pulsation and its effects on the shock-induced 
mass loss. After that, we present evolution models
of He cores with 40 - 64 $M_{\odot}$. We examine their
properties from four aspects, the thermodynamic, mass loss,
energetics and chemical properties. 

In Section \ref{sec:SLSN} we examine
the connections of our models to 
super-luminous supernova progenitors. 

In Section \ref{sec:BH} we compare the final stellar
mass of our PPISN models with the recently measured black hole masses
detected by gravitational wave signals.

In Section \ref{sec:conclusion} we conclude our results. 

We present in the Appendix
\ref{sec:comparison} 
our numerical models with those in the literature.
In Appendix 
\ref{sec:conv_mix} and Appendix \ref{sec:art_vis}
the effects of some physical inputs in the numerical 
modeling, including  the convective mixing and
artificial viscosity.

\section{Methods}
\label{sec:methods}

\subsection{Stellar Evolution}
To prepare the pre-collapse model, we use the open source
code Modules for Experiments in Stellar Astrophysics (MESA) (v8118; \cite{Paxton2011, Paxton2013, Paxton2015,Paxton2017}). 
It is a one-dimensional stellar 
evolution code.
Recent updates of this code have
also included packages for stellar pulsation analysis and 
implicit hydrodynamics extension with artificial viscosity.
We modify the package {\it ccsn} to build a He-core
or main-sequence star models directly and then we switch 
to the hydrodynamics formalism according to the global
dynamical timescale of the star.

\subsection{Hydrodynamics}

To understand the behaviour of pulsation and runaway burning in
the O-core, we use
the one-dimensional implicit hydrodynamics option.
This option appears in the third
instrument paper \citep{Paxton2015}. 
The energy conserving scheme, coupled with the 
implicit mass-conserving property of the Lagrangian formalism,
allows us to trace the evolution of the star consistently. 

We refer the readers to the instrument paper 
\cite{Paxton2015} (Section 4) where the detailed implementation
of this mass- and energy-conserving implicit
hydrodynamics scheme is documented. Here
we briefly outline the specific points
which are relevant to our calculation here. 

The realization of this scheme relies on the 
use of artificial viscosity as a substitute
to the exact Riemann solver. 
To capture the shock, the artificial viscosity
takes the form
\begin{equation}
Q_i = -C_{{\rm a}} \rho_i \frac{4 \pi r_i^6}{dm_i} \left( \frac{v_{i+1}}{r_{i+1}} - \frac{v_i}{r_i} \right),
\end{equation}
which has the same unit as the pressure term and 
it enters the system of equations by 
$P \rightarrow (P + Q)$. 
$v_i$ and $r_i$ are the velocity and radius of 
the mass shell $i$ defined at the cell boundary. 
$dm_i$ is the mass of the fluid element. 
We choose $C_{\rm a} = 0.002 - 0.02$.
However, the value of $C_{\rm a}$ is needed to be chosen by experience. 
A too large $C_{\rm a}$ may dissipate too 
early the propagation of outgoing waves, which artificially
suppresses the mass loss. A too small $C_{\rm a}$ may create
numerical difficulties when the shock becomes too strong
for the hydrodynamics to handle, especially near the surface.
We study the effects on the choice of $C_{\rm a}$
in the Appendix \ref{sec:art_vis}.

We define the physical quantities of convention as follows. 
Density, temperature, isotope mass fractions, 
specific internal and related thermodynamics quantities
are defined at the cell centers. Position, velocity,
acceleration and gravity source terms are defined at
the cell boundaries. We impose the innermost
boundary conditions as $r_0 = 0$.


The typical timescale during the pulsation is comparable to the dynamical timescale. 
However, after the pulsation phase, it is the 
Kelvin-Helmholtz (KH) timescale that dominates
the contraction. 
Even with the implicit nature of the dynamics code,
simply using the hydrodynamics formalism
to evolve the whole pulsation phase is computationally 
challenging as the Courant-Friedrich-Levy condition 
limits the maximum possible timescale,
despite the virtue of consistency in 
our calculation. We set conditions for the code to 
switch back to the hydrostatic approximation.
When the star sufficiently expands after 
bounce so that the evolutionary timescale is dominated by thermal timescale, 
we increase the maximum timestep at every 100 
steps. When the star can evolve continuously with the maximum 
timestep ($10^5$ times of Courant timestep), 
we change to the hydrostatic approximation 
to evolve the star until another pulsation
starts. If the star appears 
to be non-static during the 100-step buffer, the 
buffer is extended until the star is fully relaxed. 
The convective mixing is also switched on only in the hydrostatic mode.
In general, we find that dynamical treatment is necessary when
the central temperature of the star exceeds 
$10^{9.3}$ K. 

In the pulsation phase, once the expansion of the star
reaches the surface, it develops into a high velocity outburst
due to the density gradient near the surface. The fluid 
elements can have a velocity larger than the escape velocity. 
The ejected mass is dynamically irrelevant to the core evolution.
We remove from those mass elements which satisfy 
this condition and have a density below $10^{-6}$ g cm$^{-3}$.
We set a mass loss rate according to how fast the outermost shell
leaves our system according to its velocity. To avoid removing
mass shell in an unphysical rate due to interpolation, the
mass loss rate is capped above.

\subsection{Microphysics}

The code uses the Helmholtz equation of state \citep{Timmes1999},
which contains electron gas with arbitrary relativistic and
degeneracy levels, ions in the form of an classical ideal gas, 
photon gas with Planck distribution and electron-positron 
pairs. To model the nuclear reactions, we use the 
{\it 'approx21$\_$plus$\_$co56.net'} network. This includes the 
$\alpha$-chain network ($^{4}$He, $^{12}$C, $^{16}$O, 
$^{20}$Ne, $^{24}$Mg, $^{28}$Si, $^{32}$S, $^{36}$Ar, 
$^{40}$Ca, $^{44}$Ti, $^{48}$Cr, $^{52}$Fe and $^{56}$Ni), 
$^{1}$H, $^{3}$He and $^{14}$N for the hydrogen burning
and CNO cycle, and $^{56}$Fe and $^{56}$Co to trace
the decay chain of $^{56}$Ni. $^{56}$Cr is included 
to mimic the neutron-rich isotopes formed after 
electron capture in nuclear statistical equilibrium (NSE).

The MESA EOS is a blend of the OPAL \citep{Rogers2002}, SCVH
\citep{Saumon1995}, PTEH \citep{Pols1995}, HELM
\citep{Timmes2000}, and PC \citep{Potekhin2010} EOSes.

Radiative opacities are primarily from OPAL \citep{Iglesias1993,
Iglesias1996}, with low-temperature data from \citep{Ferguson2005}
and the high-temperature, Compton-scattering dominated regime by
\citep{Buchler1976}.  Electron conduction opacities are from
\citep{Cassisi2007}.

Nuclear reaction rates are from JINA REACLIB \citep{Cyburt2010} plus additional
tabulated weak reaction rates \citep{Fuller1985, Oda1994, Langanke2000}.
Screening is included via the prescription of \citep{Salpeter1954,Dewitt1973,Alastuey1978,Itoh1979}.
Thermal neutrino loss rates are from \cite{Itoh1996a}.

\subsection{Convective Mixing}

As indicated in \cite{Woosley2017}, convective mixing
is important that it redistributes the 
fuel and ash in the remnant core. This affects the 
subsequent nuclear burning when the star contracts 
again. We choose the Mixing Length Theory (MLT) \citep{Boehm1958} (See
e.g. \cite{Cox1968} for a realization) to model 
the convective process with Schwarzschild criterion. The MLT approximation is used
in the main-sequence phase and also when the star enters
the expansion phase. 

We have attempted to couple the convective 
mixing in the dynamical phase but it results in numerical instabilities.
We notice that the convective timescale during the 
dynamical phase is longer than dynamical timescale.
Furthermore, the more massive the star is, the burning timescale
and its contraction timescale due to PPI decrease. 
The mixing process becomes more inefficient compared to the lower mass stars.
Since we are interested to the mass loss process of PPISN, 
this means to a good approximation by neglecting convection.
(Also see Section \ref{sec:hydro} the corresponding Kippenhahn
Diagrams). Therefore, it becomes numerically manageable while
physically consistent to ignore convective mixing in the dynamical
phase. 

However, we also notice that in the lower mass regime 
(e.g. 40 $M_{\odot}$ case), the contraction timescale is on the contrary
long that the convective mixing becomes more important. In the Appendix
we examine the importance of mixing to the pulsation history for the
lower mass PPISN. We remind that during pulsations not 
only convective mixing is suppressed, but also convective energy 
transport is suppressed. This plays a major role in the weak 
pulsations found by other authors, as the ones shown in the 
Models He36 and He40 in Figure 3 of \cite{Woosley2017}. 
In these cases there is only a mild collapse, and the energy 
produced through nuclear burning can be transported 
by convection without an eruption.

\section{Evolution of PPISN Progenitors}
\label{sec:MainSeq}

In this section we cover the methodology and the results for
the stellar evolution model based on H main-sequence stars and run until 
the central temperature reaches $10^{9.4}$ K\footnote{We uploaded the related 
configuration files used in our simulations in DOI:10.5281/zenodo.3457295}.  
We choose
the \textit{Dutch} mass loss rate (an ensemble of mass loss rates
computed separately in \cite{Vink2001, Glebbeek2009}) 
for hot hydrogen rich stars, 
\cite{Nugis2000} for hot hydrogen poor stars, 
and \cite{deJager1988} for winds from cold stars. 
The scaling factor follows \cite{Maeder2001} for modeling non-rotating stars.

\begin{table*}
\begin{center}
\label{table:He_MS_Omega0}
\caption{The pre-pulsation He core mass at the exhaustion of H in the 
core. The numbers in brackets are the CO core mass at the exhaustion
of He in the core. All masses are in units of solar mass. 
\newline
$^1$ The models assume no mass loss.}
\begin{tabular}{|c|c|c|c|c|c|c|}
\hline
Mass $(M_{\odot})$ 	& $Z = 10^{-3} ~Z_{\odot} ~^1$ & $Z = 10^{-2} ~Z_{\odot} ~^1$	   & $Z = 0.1 ~Z_{\odot}$ &
$Z = 0.5 ~Z_{\odot}$	& $Z = ~0.75 Z_{\odot}$	& $Z = 1 ~Z_{\odot}$		 	\\ \hline
80 &  34.05 & 37.40 (27.20) & 33.80 (23.93) & 30.10 (23.96) & 23.60 (21.09) & 22.70 (18.66) \\
100 & 44.51 & 49.44 (37.69) & 47.16 (34.22) & 33.00 (30.65) & 31.70 (28.50) & 30.30 (24.85) \\
120 & 54.87 & 64.71 (48.40) & 59.95 (43.48) & 57.10 (41.20) & 37.40 (31.73) & 15.50 (12.02) \\
140 & 65.87 & nil           & 70.85 (56.67) & 60.78 (50.36) & 20.80 (16.90) & 12.80 (9.60) \\
160 & 76.50 & 83.31 (82.40) & 89.99 (89.12) & 52.93 (46.46) & 15.00 (11.63) & 11.99 (8.90) \\ \hline
\end{tabular}
\end{center}
\end{table*}

\subsection{Evolution in  Kippenhahn Diagram}

\begin{figure}
\centering
\includegraphics*[width=8cm,height=5.7cm]{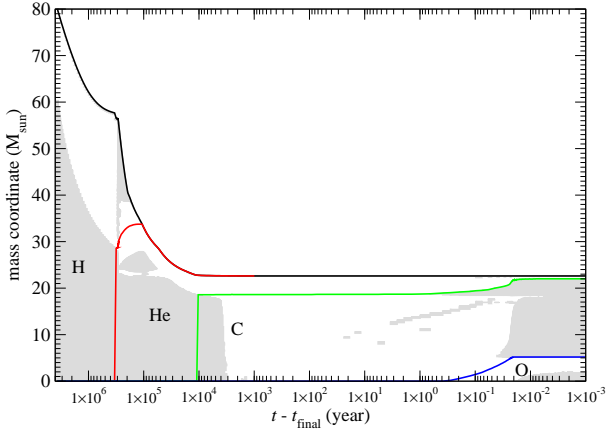}
\caption{Kippenhahn diagram of the main-sequence 
of $M = 80 ~M_{\odot}$ at solar metallicity from
the H-burning. The time stands
for the time before the core reaches a temperature of $10^{9.4}$ K.}
\label{fig:LOGS_M80_Z002_KD_plot}
\end{figure}

\begin{figure}
\centering
\includegraphics*[width=8cm,height=5.7cm]{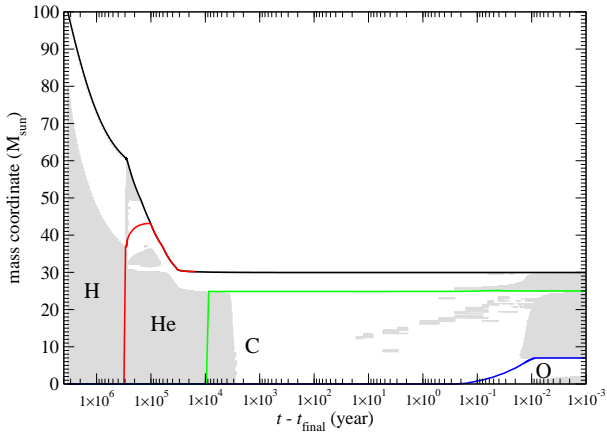}
\caption{Similar to Figure \ref{fig:LOGS_M80_Z002_KD_plot},
but for $M = 100 ~M_{\odot}$.}
\label{fig:LOGS_M100_Z002_KD_plot}
\end{figure}

\begin{figure}
\centering
\includegraphics*[width=8cm,height=5.7cm]{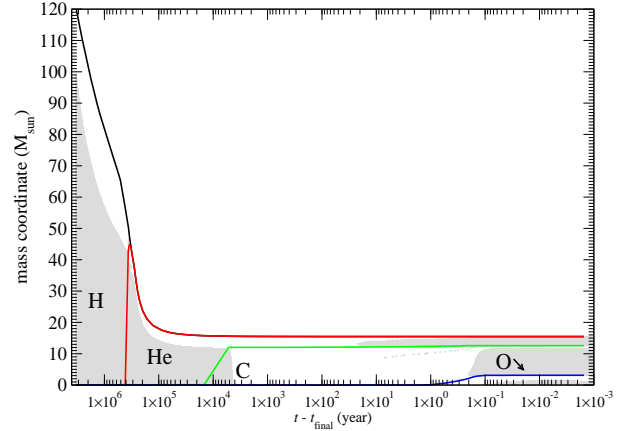}
\caption{Similar to Figure \ref{fig:LOGS_M80_Z002_KD_plot},
but for $M = 120 ~M_{\odot}$.}
\label{fig:LOGS_M120_Z002_KD_plot}
\end{figure}

In Figures \ref{fig:LOGS_M80_Z002_KD_plot}, \ref{fig:LOGS_M100_Z002_KD_plot}
and \ref{fig:LOGS_M120_Z002_KD_plot}, we plot the Kippenhahn Diagram 
of the stars with $M = 80, ~100$ and 120 $M_{\odot}$ at $Z = Z_{\odot}$.
The lines (red, green and blue) correspond to the He-,
C- and O- core mass coordinate respectively. Grey shaded regions
are the convective zones inside the star. All models are run
until the core reaches a central temperature of $10^{9.4}$ K.

At solar metallicity
the high metal content in the initial composition has largely 
increased the opacity, which allows strong mass loss during 
H-burning and He-burning due to its intrinsic high luminosity. 
It has an extremely large mass loss rate that half of the matter is 
lost in the helium burning phase for $M = $ 80 and 100 $M_{\odot}$.
and in the hydrogen burning phase for 120 $M_{\odot}$. The whole 
H envelope is lost during He-burning, which occurs about 
$10^5$ year before collapse. The initial He core mass can 
reach about half of the initial mass, but it gradually decreases
due to the later mass loss. Also, in all
three models, after the removal of H-envelope or He-burning, 
the C-core quickly forms with a mass similar to the He-core mass.
For 80 and 100 $M_{\odot}$ models they have a C core mass $\sim$ 20 $M_{\odot}$
which remains unchanged after it has been formed. The 120 $M_{\odot}$
one has a somewhat smaller one due to the previous drastic mass loss. 

The convective pattern of the star is consistent with typical massive stars. 
In H-burning phase, the core is mostly convective, while the surface is
radiative. In the He-burning phase, the core remains convective while 
some H-envelope becomes convective. But this feature disappears when
the mass loss sheds away the H-envelope. Once the C-core has formed
at about $10^4$ year before pulsation,
the star begins to contract rapidly. The core becomes radiative.
But together with the C-core and C-envelope burning, layers of 
convective shells appear. They gradually propagates and reach the C-core 
surface. When the core starts O-burning ($10^{-1}$ year from
the onset of first pulsation), the strong energy generation triggers
large scale convection that the whole C-envelope becomes convective. 
The inner core of O-rich region also becomes convective. 

\begin{figure}
\centering
\includegraphics*[width=8cm,height=5.7cm]{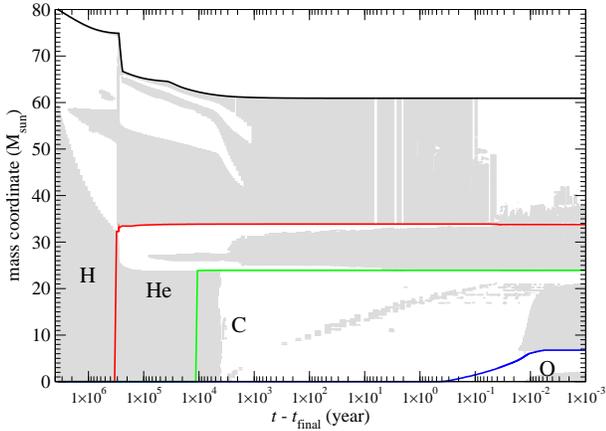}
\caption{Kippenhahn diagram of the main-sequence 
of $M = 80 ~M_{\odot}$ at 0.1 $Z_{\odot}$ from
the H-burning until the core reaches a temperature of $10^{9.4}$ K.}
\label{fig:LOGS_M80_Z0002_KD_plot}
\end{figure}

\begin{figure}
\centering
\includegraphics*[width=8cm,height=5.7cm]{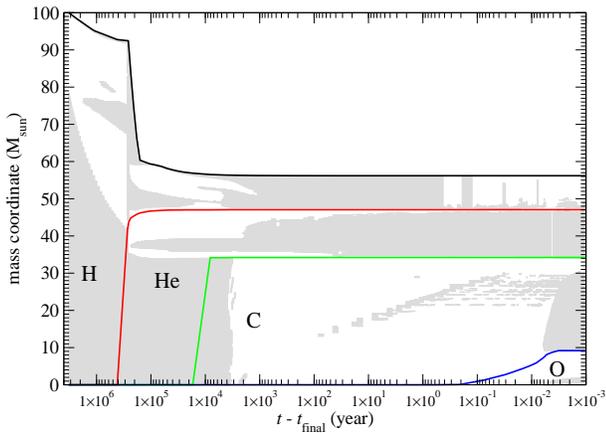}
\caption{Similar to Figure \ref{fig:LOGS_M80_Z0002_KD_plot},
but for $M = 100 ~M_{\odot}$.}
\label{fig:LOGS_M100_Z0002_KD_plot}
\end{figure}

\begin{figure}
\centering
\includegraphics*[width=8cm,height=5.7cm]{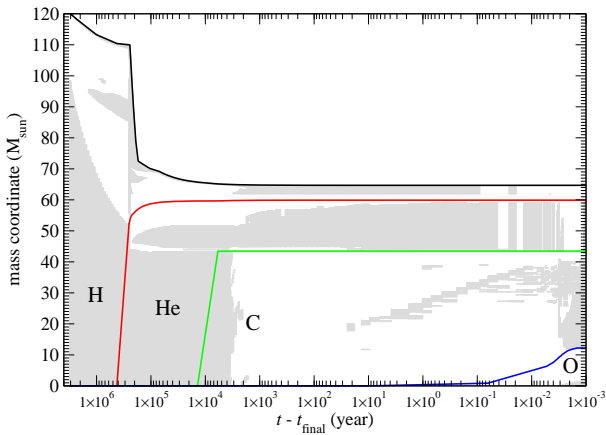}
\caption{Similar to Figure \ref{fig:LOGS_M80_Z0002_KD_plot},
but for $M = 120 ~M_{\odot}$.}
\label{fig:LOGS_M120_Z0002_KD_plot}
\end{figure}

In Figures \ref{fig:LOGS_M80_Z0002_KD_plot}, \ref{fig:LOGS_M100_Z0002_KD_plot}
and \ref{fig:LOGS_M120_Z0002_KD_plot} we plot similar to the previous
three figures but at $Z = 0.1 ~Z_{\odot}$. Different from the 
models at solar metallicity, the low metallicity implies low opacity
in the matter, and thus lowers the mass loss during the 
H- and He-burning phase. There is a clear signature of massive
He core from 30 to 50 $M_{\odot}$. 
The He core mass remains constant after it has formed. 
Near the occurrence of first pulsation,
a massive CO core $\sim$ 10 $M_{\odot}$ is also formed. 
A generally larger O-rich core is formed at the end of simulation. 

Due to the preservation of H-envelope after He-burning, the star consists
of a rich structure of convection activities before its collapse. 
The convective core has a similar structure to the higher metallicity 
case, due to the extended
H-envelope remained after He-burning, there is also a second convective zone 
which gradually move inward to the stellar core from its initial 60 $M_{\odot}$
to $\sim 40 ~M_{\odot}$. After that the He-core is fully convective during He-burning
and the convection zone extends into the H-envelope. The surface is also convective. 
During C-burning, the convective layer propagates outwards from the core to the 
C-core surface. The outer layers of He- and C- envelopes are convective. 
During contraction before the onset of O-burning, the core returns
to be radiative dominated. Similar to the high metallicity case, 
near the onset of pulsation, the core becomes convective. 

\subsection{Pre-pulsation evolution}

\begin{figure*}
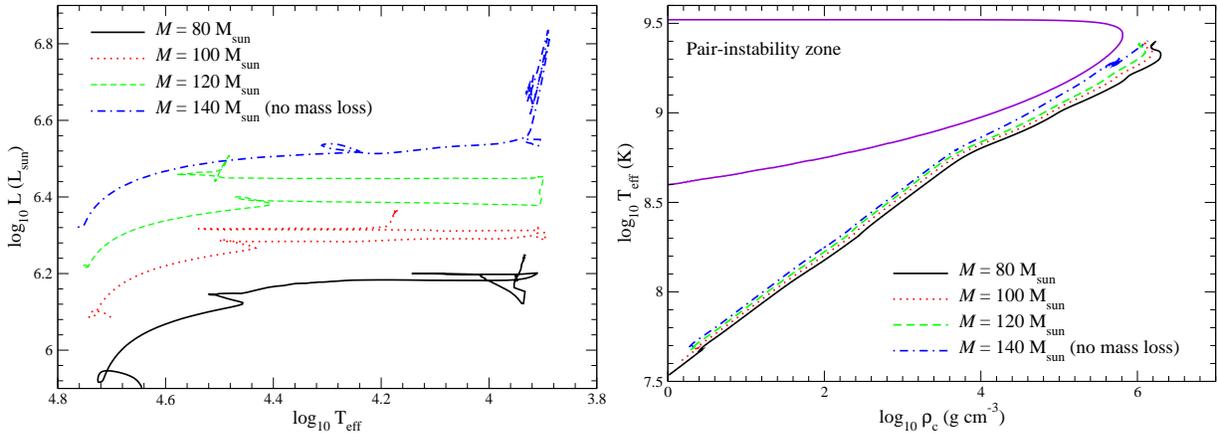

\centering
\includegraphics*[width=8cm,height=5.7cm]{HR_MS_plot.eps}
\includegraphics*[width=8cm,height=5.7cm]{tempc_rhoc_MS_plot.eps}

\caption{
(left panel) The HR diagram of the main-sequence stars from 80 to 140 
$M_{\odot}$. Notice that for the 140 $M_{\odot}$ model no mass loss
is assumed due to the later numerical instability. 
(right panel) Similar to the left panel, but for the $T_c$ against $\rho_c$
diagram. The zone enclosed by the purple curve corresponds to the instability
zone defined by $\Gamma < 4/3$.
}
\label{fig:MS_plot}
\end{figure*}

In the left panel of Figure \ref{fig:MS_plot} we plot the HR diagram 
for the main-sequence star models with $M = 80$, 100, 120 and 140 $M_{\odot}$
included. All models are fixed at $Z = 0.002 ~Z_{\odot}$. 
For numerical stability we do not include mass loss for the 140 $M_{\odot}$ model. 
In the pre-pulsation evolution, the models follow the typical HR diagram
of main-sequence star, The H-burning occurs after the star has contracted.
After H is exhausted, the star develops into red giant with He burning
which largely increases its luminosity. Depending on the mass loss,
the effective temperature can largely reduce. Also, the typical luminosity
increases with mass. 

In the right panel we plot the evolution of the central temperature
against the central density for the same set of models. 
We also draw the pair-instability zone (defined
by the adiabatic index $\Gamma < 4/3$). There is no intersection 
among models, showing that the thermodynamics properties of
the core before pulsation depend on only its mass.  
The contraction of the core follows mostly adiabatic contraction
(with a slope -3) in the diagram.

\subsection{He Core and CO Core Mass Relations}

To study the effects of metallicity and rotation, we 
perform pre-pulsation stellar evolution models
for different metallicity from $Z = 10^{-3} ~Z_{\odot}$ 
up to $Z = 1 ~Z_{\odot}$ for non-rotating main-sequence star model.
In Table \ref{table:He_MS_Omega0} we tabulate the pre-pulsation
configurations of the main-sequence stars for their
He- and CO- core masses when the core exhausts all 
H and He respectively. The CO core masses are written
in brackets. 

\begin{figure}
\centering
\includegraphics*[width=8cm,height=5.7cm]{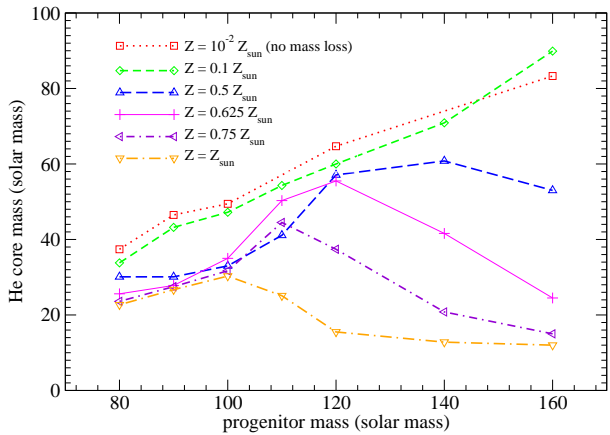}
\caption{The He core mass against progenitor mass 
when the core exhausts its hydrogen 
for stellar models at different metallicity.
For $Z = 10^{-2} ~Z_{\odot}$, the models assuming
no mass loss is assumed because of numerical 
instability encountered during He-burning in
the asymptotic red-giant branch.}
\label{fig:HeCore_plot}
\end{figure}

He core mass
grows monotonically with $M$ when $Z < 0.625 ~Z_{\odot}$.
For star models with a higher metallicity, the mass loss
rate, which is proportional to the metallicity, makes the 
He core mass drop at the high mass end. This transition starts
at a lower mass for models with a higher metallicity. 
Notice that the change and the transition mass is 
not linearly proportional to $Z$
due to the non-linear dependence of mass-loss rate. 
Also, the mass loss affects the gravity, which changes
the equilibrium structure of the star even in the 
H-burning phase. 

In Figure \ref{fig:HeCore_plot} we also plot the relations
He core mass against progenitor mass for different metallicity. 
On one hand, at low mass, the He core mass 
is not so sensitive to metallicity, that 
the He core mass approaches its asymptotic value when 
$Z \leq 10^{-2} ~Z_{\odot}$. On the other hand,
at high mass, the He core mass is very sensitive
to metallcity that from $Z = 0.625 ~Z_{\odot}$ to 
$Z = Z_{\odot}$ the He core mass can drop by 90 \%
at the star model $M = 160 ~M_{\odot}$, about 15 $M_{\odot}$.
At such low mass, the He core already leaves the 
pulsation pair-instability regime, and evolves as
a normal CCSN. Furthermore, 
the maximum He core mass for models at
solar metallicity only barely reaches the transition mass 
40 $M_{\odot}$. 

For models which completely covers the PPISN 
mass range (He star of mass 40 -- 64 $M_{\odot}$), 
we require stellar model with a metallcity at
most $0.1 ~Z_{\odot}$. 
This shows that the PPISN is very 
sensitive to the progenitor metallicity, while 
stars with solar metallicity are less likely to 
form PPISN owing to its mass loss.
 
\begin{figure}
\centering
\includegraphics*[width=8cm,height=5.7cm]{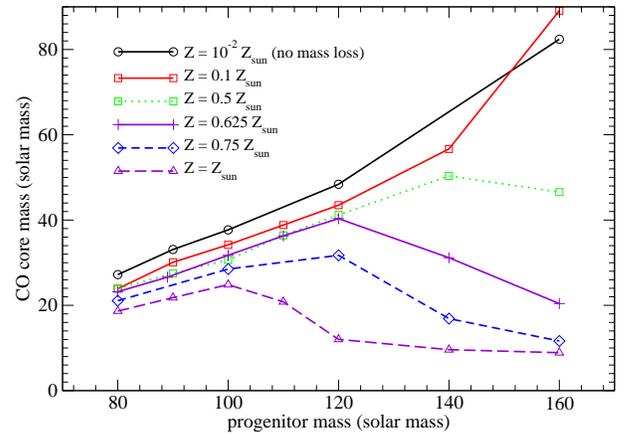}
\caption{The C core mass against progenitor mass 
when the core exhausts its He
for stellar models at different metallicity.
Again, for $Z = 10^{-2} ~Z_{\odot}$ the models
assume no mass-loss due to numerical instability.}
\label{fig:COCore_plot}
\end{figure} 

Then we examine the CO core mass. Before the CO core
mass can be defined, the massive CO core has already 
started its contraction, which increases the CO core mass. 
The metallicity effect is a similar trend to the He core mass. In all models, 
at the lower mass branch CO core mass in general increases
with progenitor mass, but it drops at the high mass
end. The CO core mass also shows a monotonic decreasing 
relation against metallicity for the same progenitor mass. 
The mass loss effect in near solar metallicity models
are more significant that the CO core mass contributes to 
less than 10 \% of the stellar mass, while those with a lower
metallicity can be about one third of the progenitor mass. 

In Figure \ref{fig:COCore_plot} we also plot the relation
CO core mass against progenitor mass at different metallicity.
The significance of the metallicity of mass loss rate can
be seen. By increasing models from 0.5 $Z_{\odot}$ to 0.75 $Z_{\odot}$, 
the CO core mass can drop by 75 \% at $M = 160 ~M_{\odot}$. 
The CO core mass shows a clearer relation than the He core mass. 
The mass scaled clearly with metallicity, except at $M = 160 ~M_{\odot}$,
where the model without mass loss ($Z = 10^{-2} ~Z_{\odot}$) 
has a lower mass than its counterpart of $Z = 0.1 ~Z_{\odot}$. 

\section{Pulsational Pair-Instability in Helium Stars}
\label{sec:hydro}

In this section we study the evolution of PPISN using the
zero-age He main-sequence as the initial condition. We 
consider stellar model with pure He, i.e. zero metallicity,
as indicated that a massive He-core is likely to be form
when $Z \leq 10^{-1} ~Z_{\odot}$.
We do not evolve dynamically
with H because
the H-envelope is not tightly bounded by the
gravitational well of the star. It is easily disturbed
and obtains high velocity during shock outbreak.
We find that to keep the H-envelope while evolving the whole star 
is computationally difficult. 
However, as the H-envelope does not couple strongly to the 
inner core, the pulsation dynamics is not 
significantly changed, when we do not consider the 
effects of H-envelope. Therefore, in this section, we consider
the dynamics, energetics, mass loss and chemical properties
of the PPISN by using the He-star as the initial condition. 
However, we notice that in general a He-star does not
always one-one correspond to the He-core evolved from traditional
stellar evolution.
We also remark that the core masses for the major elements
are defined by the mass coordinate where that particular
element (or major isotope) reaches a mass fraction $> 1$ \%.
The convective mixing is switched off when we use the hydrodynamics
option because of the numerical difficulties. In fact, the dynamical 
timescale can be shorter than mixing timescale when the shock
has formed or dynamically expanding. It is unclear for those 
scenarios whether convection can be formed robustly.
An incomplete mixing model or time-dependent convection 
model is necessary for following this part of input physics.

\begin{table*}
\begin{center}

\caption{The main-sequence star models prepared by the MESA code.
$M_{{\rm ini}}$ and $M_{{\rm fin}}$ are the initial and final 
masses of the star. $M_{{\rm H}}$, $M_{{\rm He}}$, $M_{{\rm CO}}$
are the hydrogen, He- and CO-mass before the
hydrodynamics code starts. All masses are in units of solar mass.}
\label{table:He_MS_result}
\begin{tabular}{|c|c|c|c|c|c|c|c|}
\hline

Model & $M_{{\rm ini}}$ & $M_{{\rm fin}}$ & $M_{{\rm H}}$ & $M_{{\rm He}}$ & $M_{{\rm C}}$ & $M_{{\rm O}}$ & remarks \\ \hline
He40A & 40 & 40 & 0 & 6.79 & 3.13 & 27.5 & only He core \\
He45A & 45 & 45 & 0 & 7.38 & 4.03 & 31.3 & only He core \\
He50A & 50 & 50 & 0 & 7.82 & 4.16 & 35.2 & only He core \\
He55A & 55 & 55 & 0 & 8.27 & 4.30 & 39.0 & only He core \\
He60A & 60 & 60 & 0 & 8.69 & 4.43 & 42.9 & only He core \\
He62A & 62 & 62 & 0 & 8.77 & 4.59 & 44.6 & only He core \\
63HeA & 63 & 63 & 0 & 8.89 & 4.64 & 45.3 & only He core \\
He64A & 64 & 64 & 0 & 8.96 & 4.63 & 46.1 & only He core \\ \hline

\end{tabular}
\end{center}
\end{table*}

\subsection{Evolution in Kippenhahn Diagram}
In this part we examine the overall evolution of the
PPISN evolved from He core until the onset of Fe-core collapse.

\begin{figure*}
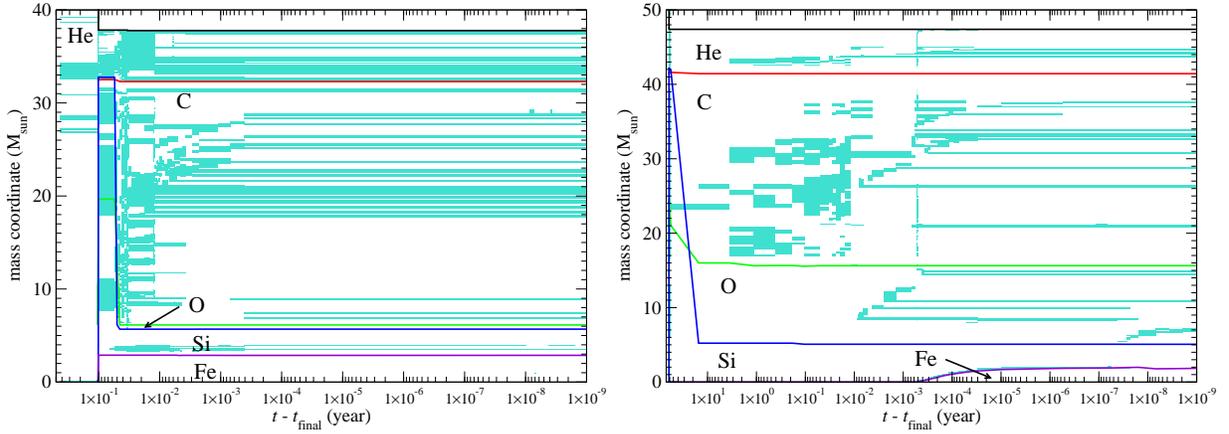

\centering
\includegraphics*[width=8cm,height=5.7cm]{fig9.eps}
\includegraphics*[width=8cm,height=5.7cm]{fig10.eps}
\caption{(left panel) Kipperhahn Diagram for the Model He40A 
($M_{{\rm He}} = 40 ~M_{\odot}$) until the onset of final
collapse. The lines correspond to the inner boundary
where the mass fractions of the respective elements 
drop below $10^{-2}$. By this definition, the surface mass
coordinate of
the star, if it does not experience strong mass ejection,
is the He-core mass since we start from a He-star.
(right panel) Similar to Figure \ref{fig:LOGS_He40_KD_plot},
but for $M_{{\rm He}} = 50 ~M_{\odot}$. 
The time on the x-axis is defined by the time before the onset of final collapse.}
\label{fig:LOGS_He40_KD_plot}
\end{figure*}

\begin{figure*}
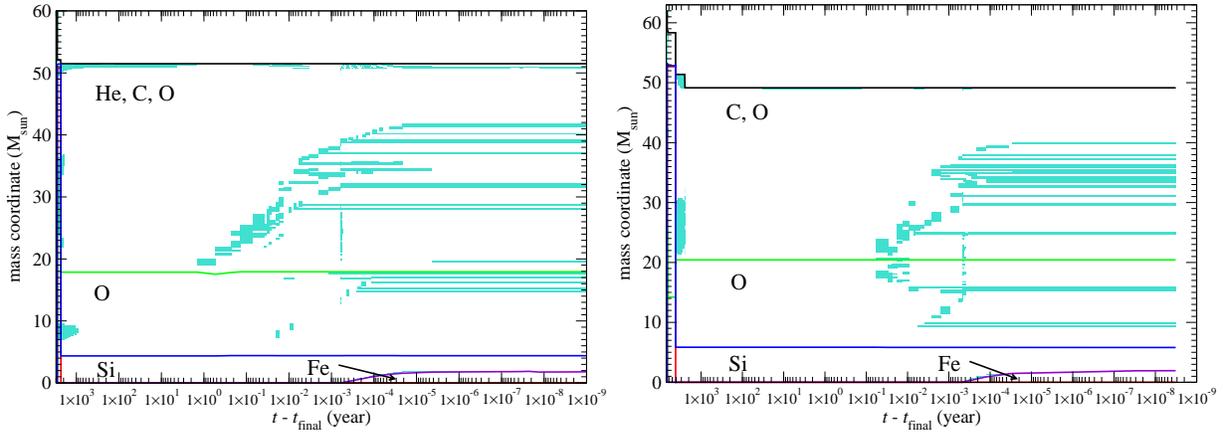

\centering
\includegraphics*[width=8cm,height=5.7cm]{fig11.eps}
\includegraphics*[width=8cm,height=5.7cm]{fig12.eps}
\caption{(left panel) Similar to Figure \ref{fig:LOGS_He40_KD_plot},
but for $M_{{\rm He}} = 60 ~M_{\odot}$.
(right panel) Similar to Figure \ref{fig:LOGS_He40_KD_plot},
but for $M_{{\rm He}} = 62 ~M_{\odot}$.}
\label{fig:LOGS_He62_KD_plot}
\end{figure*}

In Figures \ref{fig:LOGS_He40_KD_plot} and
\ref{fig:LOGS_He62_KD_plot} we plot
the Kippenhahn diagram of Models He40A, He50A, He60A and He62A.
The coloured zone is again the convective zone while the lines
(solid, dotted, dashed, long-dash, dot-dash) are the He-, C-, O-,
Si- and Fe-core mass coordinate. The x-axis is the time counting
backward from its collapse. We define the core boundary to be 
the inner boundary of the mass fraction for that corresponding element 
to drop below $10^{-2}$. Therefore, since we start from a He core,
the He-rich surface, which is also the total mass of the star,
stands for the He-core. Notice that for the cases with strong 
mass ejection, the whole He-rich surface can be shredded off. 
Here the time is defined by the remaining time
from the onset of final collapse.

For Model He40A, after the strong pulsation, the star expands and the 
outer part of the star above 18 $M_{\odot}$
becomes convective. Also, the star established its 
O- and Si-cores at $m(r) \sim 5 ~M_{\odot}$ and its Fe-core
at $m(r) \sim 2 ~M_{\odot}$. Radiative transfer 
remains the major energy transport in the core.
Thin layers of convection shells can be found in most
parts of the He- and C-envelope. 
At about $10^{-1}$ year, the Si and O core can 
reach as far as $\sim ~30 M_{\odot}$. This is because during
the propagation of the acoustic wave near surface, the density
gradient accelerates the wave into a shock, which heats
up the matter around there. As a result, in such He-rich 
material, it facilitates the He-burning and gives product
including C, O, Ne and Si. However, accompanied with the 
extended convection during the expansion-contraction
phase, the outer O- and Si-rich zones disappear and 
the values correspond to the inner layers, which come from 
previous hydrostatic burning. 

For Model He50A, after the pulsation, the C-, O- and Si-cores
are produced simultaneously. But the O- and Si-cores quickly
retreat from 30 $M_{\odot}$ to 5 and 10 $M_{\odot}$
respectively. The early formation of O- and Si-cores is
because when the shock reaches the surface, the shock 
heating is capable in producing O- and Si-rich material 
around that region. However, away from the shock-heated
zone, no significant O- and Si-productions take place. However, 
after the production, the mixing and mass loss by
pulsation quickly remove these material. As a result,
the O- and Si-core mass coordinates return to the 
corresponding inner values, where the real O- and Si-core
locate. At $10^{-3}$ year before the final collapse,
the contraction of star allows the central density 
to be high enough for burning until NSE. The 
Fe- and neutron-rich core forms almost simultaneously
at $\sim 2 ~M_{\odot}$.
Different from Model He40A, the inner core is
no longer convective, except 
after pulsation. There is an extended period of time
at $\sim 10^{-2}$ year before its final collapse,
the star continues to hold fragmented convective layers. 

For Model He60A, there is also no inner convective
core after its pulsation. Again, the shock heating creates
a temporary outer O- and Si-core outside surface,
but they return to the inner ones after mass ejection
and mixing, to $\sim$ 5 and 15 $M_{\odot}$.
Different from the previous two models, the expanded
star after pulsation does not reach any convective
state before its second pulse or final collapse. 
An outward propagating convective structure can be
seen from $\sim 1$ year before collapse. It moves 
from $m(r) = 20 ~M_{\odot}$ -- 40 $M_{\odot}$.  
The convection zone is small that it does not contribute
in bringing the fuel from outer layer to the actively
burning layer. Similar to Model He50A, the Fe- and 
neutron-rich cores appear at $\sim 2 ~M_{\odot}$
at $10^{-3}$ year before collapse. 

For Model He62A, it is different from the previous three
models because of its extensive mass loss after pulsation. 
After the first pulse, the star reaches a very extended
period $\sim 10^4$ years of fully convective state.
Again, the convection washes away the external C- and 
O-envelope.
The first pulse creates a final C-, O-cores which
locate at 20 and 5 $M_{\odot}$ respectively. In the second pulsation,
the Fe-core is also produced which has a mass $\sim$3 $M_{\odot}$.
During its contraction at 1 year before its final collapse,
the core reaches the third convective state. 
During contraction, the outer extended convective zone
also moves outward from 20 to 40 $M_{\odot}$.
The convective structure is again fragmented.  
A 2 $M_{\odot}$ Fe-core is formed only near $10^{-3}$
year before the final collapse. 

\begin{figure*}
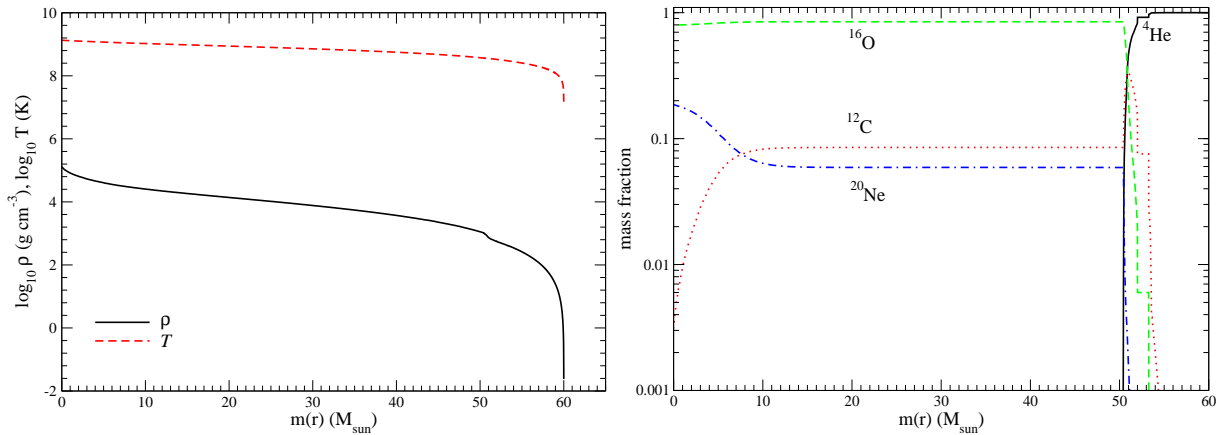

\centering
\includegraphics*[width=8cm,height=5.7cm]{He60_init_profile_plot.eps}
\includegraphics*[width=8cm,height=5.7cm]{He60_init_profile_plot2.eps}

\caption{
(left panel) The initial profile of density and temperature of Model He60A.
(right panel) Similar to the left panel, but for the chemical composition
including $^{4}$He, $^{12}$C, $^{16}$O and $^{20}$Ne. 
}
\label{fig:MS_plot2}
\end{figure*}

\subsection{Pre-Pulsation Evolution}

We first present the results for the pre-collapse profile based on the
He main-sequence star in Table \ref{table:He_MS_result}.
The pre-pulsation evolution uses the hydrostatic approximation
and it is done until the central temperature reaches $10^{9.4}$ K,
where the dynamical timescale begins to be comparable with 
the O-burning timescale. 
Below $10^{9.3}$ K the star evolves in
a quasi-static manner but not assuming hydrostatic equilibrium.
From the table we can see that
the initial He core mass affects the pre-pulsation C- and 
O- core. We choose the He-core models with a mass from 40 to 64 $M_{\odot}$,
which produce CO cores from 30.82 to 50.42 $M_{\odot}$,
with the remaining unburnt He in the envelope.

In Figure \ref{fig:MS_plot2} we present the initial model profile
and its composition. We find that most models are very similar
with each other, so for demonstration we picked $M_{{\rm He}} = 60 ~M_\odot$
as an example. The star consists of three parts: A slowly varying core which extends
up to 50 $M_{\odot}$, an envelope of rapidly falling density and 
a surface with rapidly falling temperature.

In the right panel we plot the chemical abundance profile
for the same model. 
The model contains a flat core of mostly $^{16}$O up to 
$\sim 50 ~M_{\odot}$.
Then it becomes C-rich and then He-rich until the 
surface of the star.

\begin{figure*}
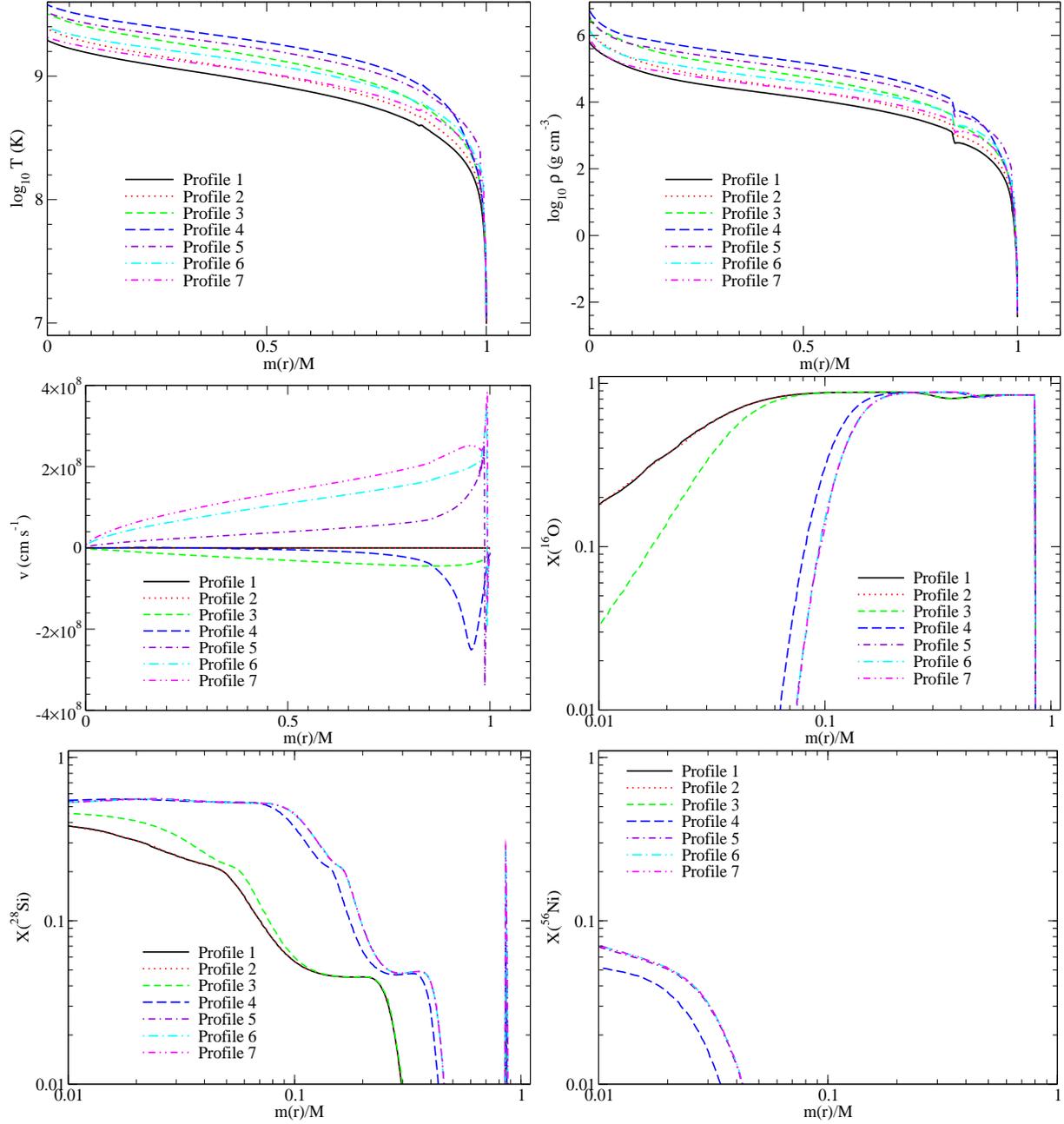

\centering
\includegraphics*[width=8cm,height=5.7cm]{fig14a.eps}
\includegraphics*[width=8cm,height=5.7cm]{fig14b.eps}
\includegraphics*[width=8cm,height=5.7cm]{fig14c.eps}
\includegraphics*[width=8cm,height=5.7cm]{fig14d.eps}
\includegraphics*[width=8cm,height=5.7cm]{fig14e.eps}
\includegraphics*[width=8cm,height=5.7cm]{fig14f.eps}

\caption{
(top left) The temperature evolution of Model He60A around the 
second pulse. The profiles are chosen in the way that the central temperature
reaches $10^{9.3}$ (Profile 1), $10^{9.4}$ (Profile 2), $10^{9.5}$ (Profile 3)
before the pulse, during the peak (Profile 4), and the central temperature
returns to $10^{9.5}$ (Profile 5), $10^{9.4}$ (Profile 6) 
and $10^{9.3}$ K (Profile 7) respectively. 
(top right) Similar to the upper left panel, but for the density profiles.
(middle left) Similar to the upper left panel, but for the velocity profiles.
(middle right) Similar to the upper left panel, but for $^{16}$O mass fraction.
(bottom left) Similar to the upper left panel, but for $^{28}$Si mass fraction.
(bottom right) Similar to the upper left panel, but for $^{56}$Ni mass fraction.}
\label{fig:pulse_plot}
\end{figure*}

\subsection{Pulsation}

We first study the time evolution of the pulsation. To do so,
we examine the second pulse of Model He60A, which is 
a strong pulse (with mass ejection) of mass
$\approx 10 ~M_{\odot}$. We choose this particular pulse
because it is strong enough to create global change
in the profile so that we can understand the changes during 
the contraction (before maximum of central temperature
in the pulsation) and expansion (after
minimum of that in the pulsation) phases. 

The core is mostly supported by the radiation
pressure. With the catastrophe in pair production, 
the supporting pressure suddenly drops, where the core
softens with corresponding equation of state adiabatic 
index $\gamma < 4/3$ in the core. 
However, unlike the stars with a mass 10 -- 80
$M_{\odot}$ which have rich Fe-cores at the moment of
their collapse, in PPISN and PISN the core is mostly made of $^{16}$O 
when contraction starts. The softened core allows a very
strong contraction and the $^{16}$O-rich core can  
reach the explosive temperature which 
releases a large amount of energy, sufficient to disrupt the star.
$^{28}$Si and $^{56}$Ni can be produced during
the contraction, where the central temperature can reach
beyond $10^{9.5}$ K. As a result, the star 
stops its contraction and expands.
The rapid expansion causes strong compression
to the matter on the surface, which efficiently
causes ejection of high velocity matter on
the surface and dissipates the energy. After that, 
the core becomes bounded again.
The pulsation restarts after it has lost 
most of its previously produced energy by radiation and neutrinos. 
The whole process repeats until the $^{56}$Fe core, formerly $^{56}$Ni, 
exceeds the Chandrasekhar mass that it collapses by its own gravity 
before the compression heating can reach the further outgoing 
$^{16}$O-rich envelope.

In Figure \ref{fig:pulse_plot} we plot in 
the top left, top right and middle left panels
the temperature, density and velocity evolution
at selected time respectively. We pick the profiles when the 
core temperature reaches $10^{9.3}$, $10^{9.4}$
and $10^{9.5}$ K before the core reaches its
peak temperature during the pulse for Profiles 1-3, at its peak 
temperature for Profile 4, and after the core has reached its
peak temperature for Profile 5 -- 7 for the same
central temperature interval. In the middle right, 
bottom left and bottom right panels we plot the chemical
abundance profiles for isotopes $^{16}$O, 
$^{28}$Si and $^{56}$Ni respectively.  

First we study the hydrodynamics quantities. 
For the temperature, in the contraction (expansion) phase
the star shows a global heating (cooling) due to the 
compression (expansion) of matter, and no 
temperature discontinuity can be observed. This 
shows that the whole star contracts adiabatically, without
producing explosive burning in the star. By comparing the temperature profiles
at the same central temperature (Profiles 1 and 7 for $T_c = 10^{9.3}$ K,
profiles 2 and 6 for $T_c = 10^{9.4}$ K and profiles
3 and 5 for $T_c = 10^{9.5}$ K), the net effect of nuclear
burning can be extracted. The part outside $q \sim 0.3$
has a higher temperature after the pulse. Similar comparison can be 
carried out for the density profile. The inner core
within $q \sim 0.3$ is unchanged after pulsation, while
the density in the outer part increases. The velocity 
profiles show more features during the pulse. Before
the star reaches its maximally compressed state, the
velocity everywhere is much less than $10^8$ cm s$^{-1}$.
At the peak of the pulse, the envelope has the 
highest infall velocity of $\sim 2 \times 10^8$ cm s$^{-1}$. 
After that, in Profile 5, the core starts the homologous expansion phase, 
with a sharp velocity discontinuity peak near the surface
between the outward going core matter and the infalling
envelope. Beyond Profile 6, the discontinuity reaches the 
surface and creates a shock breakout. The surface matter 
can freely escape from the star. 

For the chemical composition, the effects of the 
pulse becomes clear. Since the second pulse, 
part of the core $^{16}$O is already consumed in 
the first pulse, which is converted to $^{28}$Si
already. During the compression, before the core
reaches its maximum temperature, $^{16}$O is 
significantly consumed and forming 
$^{28}$Si. When the core reaches the peak temperature, 
the O within $q \approx 0.06$ is completely
burnt, where intermediate mass elements, such
as $^{28}$Si, is produced. However, Fe-peak
elements, such as $^{56}$Ni are not yet produced. 
On the other hand, during the expansion phase, 
most O-burning ceased, making the $^{16}$O and 
$^{28}$Si unchanged after the central temperature
reaches $10^{9.4}$, while advanced burning 
still proceeds slowly to form Fe-peak elements.

\subsection{Global Properties of a Pulse}

\begin{table*}
\begin{center}
\label{table:He_PPI_result}
\caption{The masses and chemical compositions of the models.
"bounce" means the number of pulse in the chronological order,
where "E" stands for the model at the end of the simulation. 
$M_{{\rm sum}}$ is the current mass in units of solar mass. 
$M_{{\rm He}}$, $M_{{\rm C}}$, 
$M_{{\rm O}}$, $M_{{\rm Mg}}$, $M_{{\rm Si}}$, $M_{{\rm IME}}$
and $M_{{\rm Fe~group}}$ are the masses of He, C, O, Mg, Si, intermediate
mass elements and and elements of nuclear statistical equilibrium 
in the star. For weak pulse, the moment is defined by the minimum 
temperature reached between pulses. For strong pulse,  
the composition is determined when the core cools down to 
a central temperature of $10^{9.3}$ K.}
\begin{tabular}{|c|c|c|c|c|c|c|c|c|c|c|c|}
\hline
Model & bounce & $M_{{\rm sum}}$ & $M_{{\rm He}}$ & $M_{{\rm C}}$ & $M_{{\rm O}}$ & 
$M_{{\rm Mg}}$ & $M_{{\rm Si}}$ & $M_{{\rm IME}}$ & $M_{{\rm Ni}}$ & $M_{{\rm Fe~group}}$ & remark \\ \hline

He40A & 1 & 40.00 & 6.77 & 2.65 & 26.70 & 0.69 & 0.96 & 3.89 & 0.00 & 0.00 & Weak \\
He40A & 2 & 40.00 & 6.65 & 2.34 & 24.70 & 0.81 & 1.99 & 6.30 & 0.00 & 0.01 & Weak \\
He40A & 3 & 40.00 & 6.59 & 2.14 & 23.16 & 0.89 & 2.95 & 7.70 & 0.11 & 0.40 & Weak \\
He40A & 4 & 40.00 & 6.57 & 2.07 & 22.52 & 0.91 & 3.07 & 7.83 & 0.24 & 1.01 & Weak \\
He40A & 5 & 40.00 & 6.54 & 1.94 & 21.77 & 0.92 & 3.13 & 7.79 & 0.24 & 1.91 & Weak \\
He40A & 6 & 40.00 & 6.51 & 1.85 & 20.06 & 0.92 & 3.43 & 7.78 & 1.69 & 3.76 & Strong \\
He40A & E & 37.78 & 4.68 & 1.79 & 20.06 & 0.91 & 3.43 & 7.77 & 0.07 & 3.42 & Final \\ \hline

He45A & 1 & 45.00 & 7.19 & 3.00 & 30.87 & 0.83 & 0.62 & 3.94 & 0.00 & 0.00 & Weak \\ 
He45A & 2 & 45.00 & 7.00 & 2.48 & 28.05 & 1.13 & 2.41 & 7.47 & 0.00 & 0.01 & Weak \\ 
He45A & 3 & 45.00 & 6.92 & 2.32 & 26.60 & 1.16 & 3.43 & 9.04 & 0.00 & 0.11 & Weak \\ 
He45A & 4 & 45.00 & 6.91 & 2.20 & 25.36 & 1.19 & 4.11 & 9.78 & 0.56 & 0.75 & Strong \\ 
He45A & E & 39.26 & 1.74 & 1.95 & 25.32 & 1.17 & 3.43 & 8.44 & 0.06 & 1.80 & Final \\ \hline

He50A & 1 & 50.00 & 7.59 & 2.95 & 34.61 & 1.10 & 0.84 & 4.85 & 0.00 & 0.00 & Weak \\
He50A & 2 & 50.00 & 7.38 & 2.41 & 31.16 & 1.37 & 3.13 & 9.03 & 0.02 & 0.02 & Weak \\
He50A & 3 & 50.00 & 7.29 & 2.15 & 28.38 & 1.35 & 5.06 & 11.62 & 0.47 & 0.57 & Strong \\
He50A & E & 47.39 & 5.21 & 2.17 & 28.38 & 1.33 & 4.08 & 9.80 & 0.09 & 1.73 & Final \\ \hline

He55A & 1 & 55.00 & 7.96 & 2.83 & 38.00 & 1.47 & 1.27 & 6.20 & 0.00 & 0.00 & Weak \\
He55A & 2 & 55.00 & 7.87 & 2.42 & 35.06 & 1.62 & 3.46 & 9.63 & 0.02 & 0.03 & Strong \\ 
He55A & 3 & 53.55 & 6.35 & 1.92 & 31.70 & 1.72 & 4.64 & 11.17 & 1.99 & 2.40 & Strong \\
He55A & E & 48.22 & 1.75 & 1.59 & 31.66 & 1.53 & 4.50 & 10.72 & 0.01 & 2.49 & Final \\ \hline

He60A & 1 & 60.00 & 8.43 & 2.75 & 41.71 & 1.72 & 1.67 & 7.11 & 0.00 & 0.00 & Strong \\
He60A & 2 & 59.52 & 7.91 & 2.22 & 36.78 & 1.64 & 5.42 & 12.46 & 0.13 & 0.15 & Strong \\ 
He60A & E & 51.48 & 0.75 & 1.92 & 36.75 & 1.61 & 4.21 & 10.32 & 0.09 & 1.64 & Final \\ \hline
          
He62A & 1 & 62.00 & 8.52 & 2.44 & 41.91 & 1.85 & 3.11 & 9.13 & 0.00 & 0.00 & Strong \\
He62A & 2 & 58.34 & 4.85 & 1.75 & 37.17 & 1.84 & 5.63 & 12.88 & 1.49 & 1.68 & Strong \\ 
He62A & E & 49.15 & 0.07 & 0.09 & 34.52 & 1.60 & 4.88 & 10.96 & 0.04 & 2.66 & Final \\ \hline
      
He64A & 1 & 64.00 & 8.69 & 2.39 & 42.87 & 1.90 & 3.63 & 10.05 & 0.00 & 0.00 & Strong \\ 
He64A & E &  0.00 & 0.00 & 0.00 &  0.00 & 0.00 & 0.00 & 0.00 & 0.00 & 0.00 & Final \\ \hline

\end{tabular}
\end{center}
\end{table*}

Here we study some representative models of He core with a mass
from 40 to 62 $M_{\odot}$. They show very different pulsation histories,
by their number of pulses and their corresponding strengths. 
In Table \ref{table:He_PPI_result} we tabulate the stellar mass
and the element mass in the star after each of the pulse. 

For Model He40A, most of the pulses are weak, however, following 
each of the pulse, mass of $^{16}$O is gradually consumed
and produce $^{28}$Si. At late pulses, where the core reaches 
beyond $10^7$ g cm$^{-3}$, NSE elements are also produced. 
In the last pulse, the core is sufficiently compressed such 
that an Fe core beyond 1.4 $M_{\odot}$ is produced,
which is followed by later mass loss. Most of the 
ejected mass is He. 

For Model He45A, most of the pulses are weak. 
With the number of pulses increases, not only Si, but also
$^{56}$Ni are produced. The last pulse, which is the strongest
overall, produces about 0.56 $M_{\odot}$ Ni, while the 
generated heat creates a shock to eject about 6 $M_{\odot}$
matter before the final collapse. 

For Model He50A, the number of pulses becomes smaller and 
again only the last pulse is a strong pulse which can 
eject mass. Compared to previous models, in each pulse 
more $^{16}$O is consumed, which produces Si. At the final
strong pulse, less Ni is produced, while the accompanying
mass loss ejects the He in the envelope. It should be noted that
its lower mass ejection compared to Model He45A comes from
the difference that the O in Model He45A is burnt in 
a much compressed state. This creates a much stronger
pulsation when the expansion approaches the surface, 
which increases the mass loss. 

For Model He55A, it has two strong pulses,
in contrary to lower mass models having only one strong pulse. Its pulses are 
qualitatively similar to Model He50A. 

For Model He60A, it has no weak pulse.
The contraction always makes a significant mass
of O to be burnt to produce the thermal pressure
to support the softened core against its
contraction. Due to the strong mass ejection, at the
end of the simulation the star almost runs out of He. 
However, one difference of this model from the others 
is that it has a much lower Ni mass after
pulsation. Most of the Fe, which leads to the collapse,
is created during the contraction towards collapse.

For Model He62A, it has a similar pulse pattern as
Model He60A but is stronger. Each pulse can 
consume about $3 - 4 ~M_{\odot}$ of O. Different 
from previous models, Model He62A has an abundant amount of 
O even during its contraction towards collapse, 
and O continues to be consumed before it collapses. 

For Model He64A, which is a pair-instability supernova  
instead of PPISN, there is only one pulse before its
total destruction. Due to its much lower density 
when large-scale O-burning occurs, even when about 
a few solar mass of O is burnt during the pulse, the
energy is sufficient to eject all mass when the 
pulse reaches the surface.

\subsection{Thermodynamics}

\begin{figure}
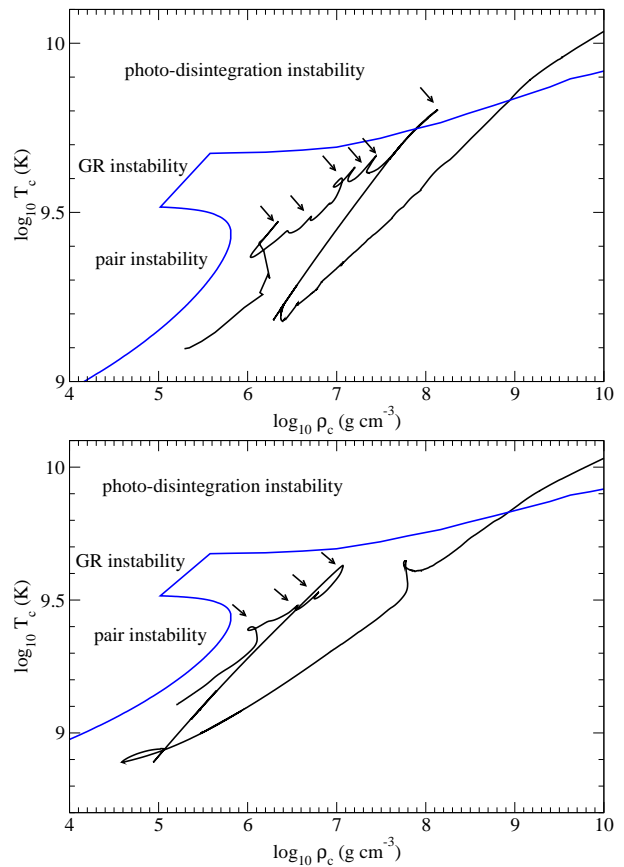

\centering
\includegraphics*[width=8cm,height=5.7cm]{fig15a.eps}
\includegraphics*[width=8cm,height=5.7cm]{fig15b.eps}
\caption{$(cont'd)$
(upper panel) The central temperature against central density for Model He40A.
(lower panel) Similar to the upper panel, but for Model He45A.
In each plot, the region on the left of the blue line 
stands for regimes dominated by the dynamical instability of pair creation, 
general relativistic effects (see, e.g., \cite{Osaki1966}) and photo-disintegration
of matter in NSE at $Y_{\rm e} = 0.5$ \citep{Ohkubo2009}.
The arrows indicate where the pulsations take place.}
\label{fig:tempc_rhoc_plot}
\end{figure}

\begin{figure}
\centering
\includegraphics*[width=8cm,height=5.7cm]{fig16a.eps}
\includegraphics*[width=8cm,height=5.7cm]{fig16b.eps}
\caption{$(cont'd)$
(upper panel) The central temperature against central density for Model He50A.
(lower panel) Similar to the upper panel, but for Model He55A.
The blue lines and the arrows follow the same meaning as in Figure \ref{fig:tempc_rhoc_plot}.}
\label{fig:tempc_rhoc_plot2}
\end{figure}

\begin{figure}
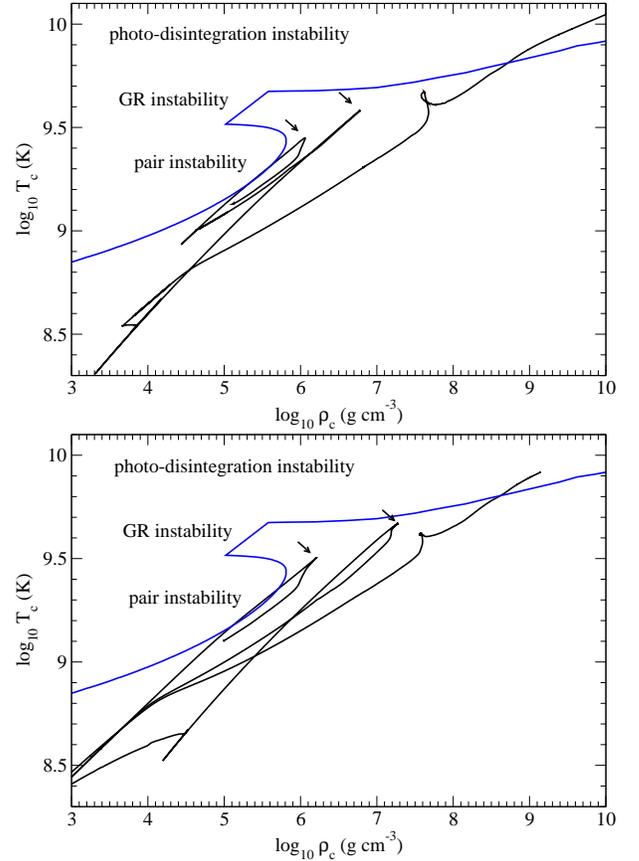

\centering
\includegraphics*[width=8cm,height=5.7cm]{fig17a.eps}
\includegraphics*[width=8cm,height=5.7cm]{fig17b.eps}

\caption{$(cont'd)$
(upper panel) The central temperature against central density for Model He60A.
(lower panel) Similar to the upper panel, but for Model He62A.
The blue lines and the arrows follow the same meaning as in Figure \ref{fig:tempc_rhoc_plot}.}
\label{fig:tempc_rhoc_plot3}
\end{figure}

In Figures \ref{fig:tempc_rhoc_plot}, \ref{fig:tempc_rhoc_plot2}
and \ref{fig:tempc_rhoc_plot3}
we plot the central density and temperature against time for Models
He40A ,He45A, He50A, He55A, He60A and He62A in the
six panels respectively. To show that the rapid contraction comes
from the PPI, we show in each plot the zones where
electron-positron pair creation, the dynamical instability 
induced by photo-disintegration of matter in NSE at $Y_{\rm e} = 0.5$
\citep{Ohkubo2009} and 
the dynamical instability induced by general relativistic effects \citep{Osaki1966}.
The arrows in the figure
show where the pulses take place. Here we define weak and strong
pulse to the pulsation of the star without or with mass ejection.
The strength of the pulse is further defined by how much the 
core expands and cools down.

For Model He40A, at the beginning the central density is 
the highest among all six models. It has thus weaker
pulses because the core is more compact and degenerate. It has 
five weak pulses and one strong pulse 
(indicated by arrows in the figure)
where each of the small pulses only leads to a small drop of the central 
density and temperature. Then the core quickly resumes its
contraction again. Only at the final pulse, when the 
core begins to reach the Fe photo-disintegration zone, 
the softened core leads to a fast contraction and
reaches a central temperature $T_c = 10^{9.8}$ K. 
This triggers a large scale O-burning in the 
outer core, which leads to a drastic drop in the central
density and temperature, showing that the star is 
expanding, until the $T_c$ reaches $10^{9.2}$ K. 
Then the core resumes its contraction.
Since most O in the core is burnt, the Si-burning 
cannot produce adequate energy to create further pulsations. The star directly collapses.

For Model He45A, it shows a fewer number of pulses than Model He40A.
It has three weak pulses and one strong pulse. The initial
path is closer to the PC instability zone. The last 
pulse is triggered at $T_c = 10^{9.7}$ K 
and has a lowest $T_c$ of $10^{8.9}$ K when
it is fully expanded. 

For Model He50A, it has only two weak pulses
and one strong pulse. The evolution shows less 
structure compared to the previous two models
because of the earlier trigger of large-scale 
O-burning in the core. The core starts the 
big pulse when $T_c = 10^{9.6}$ K and its 
expansion makes $T_c$ reaching $10^{8.8}$ K
at minimum. Before its collapse, there is a
small wiggle along its trajectory. We notice that
at this phase the core has a small pulsation
when the core becomes degenerate. 

\begin{figure*}
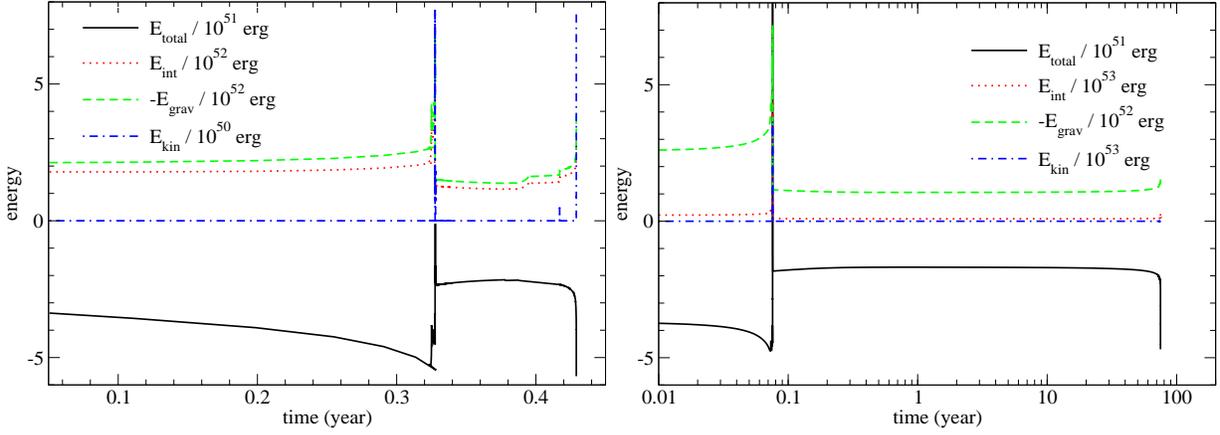

\centering
\includegraphics*[width=8cm,height=5.7cm]{fig18a.eps}
\includegraphics*[width=8cm,height=5.7cm]{fig18b.eps}
\caption{
Total $E_{{\rm total}}$, 
internal $E_{{\rm int}}$,  net gravitational $E_{{\rm grav}}$ 
and kinetic $E_{{\rm kin}}$ energies against time for
Models He40A (left panel) and He45A (right panel) respectively.}
\label{fig:energy_plot}
\end{figure*}

\begin{figure*}
\centering
\includegraphics*[width=8cm,height=5.7cm]{fig19a.eps}
\includegraphics*[width=8cm,height=5.7cm]{fig19b.eps}
\caption{
Similar to Figure \ref{fig:energy_plot}, but for 
Models He50A (left panel) and He55A (right panel) respectively.}
\label{fig:energy_plot2}
\end{figure*}

\begin{figure*}
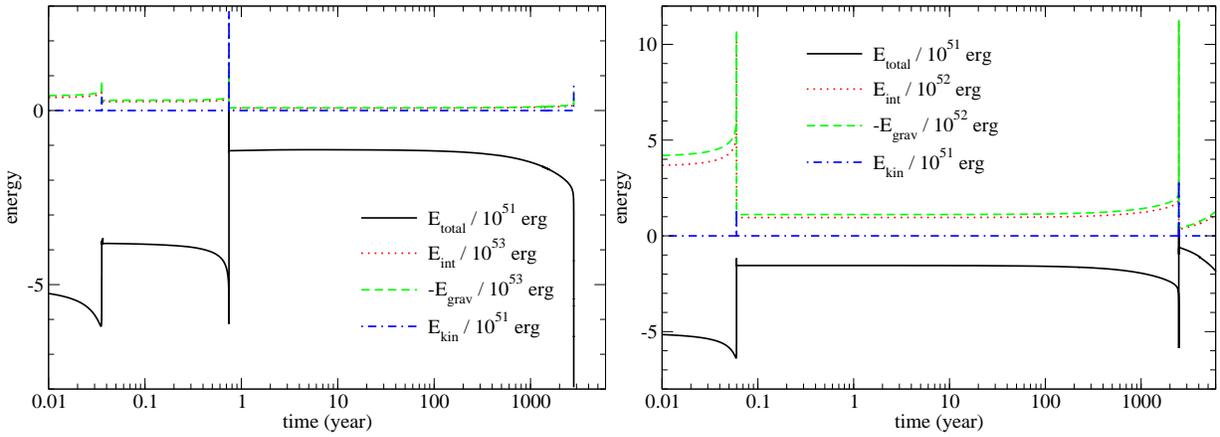

\centering

\includegraphics*[width=8cm,height=5.7cm]{fig20a.eps}
\includegraphics*[width=8cm,height=5.7cm]{fig20b.eps}

\caption{$(cont'd)$
Similar to Figure \ref{fig:energy_plot} but
for Models He60A (left panel) and He62A (right panel) respectively.}
\label{fig:energy_plot3}
\end{figure*}

For Model He55A, it has one weak pulse and two
strong pulses. The two strong pulses start when 
$T_c$ reaches $10^{9.5}$ and $10^{9.7}$ K
respectively, with a minimum temperature
after relaxation at $10^{8.8}$ and $10^{8.6}$ K.

For Model He60A, there is no weak pulse and
two strong pulses, where the stellar core intersects
with the PC instability zone during its expansion.
The two pulses start when $T_c$ reaches 
$10^{9.4}$ and $10^{9.6}$ K. The core
finishes its expansion when it reaches 
$10^{9.0}$ and $10^{8.3}$ K. 

For Model He62A, the star model becomes very close
to the PC instability where the core enters
the zone for a short period of time during its
expansion. It is similar to Model He60A that there
are two strong pulses. The two peaks start at
$10^{9.5}$ and $10^{9.7}$ K while both 
pulses end at a minimum temperature of
$10^{8.4}$ K, showing that the two pulses are
of similar strength. After that, the core 
starts collapsing similar to all other five models. 

By comparing all six models, we can observe the following 
trend for the pulse structure as a function of progenitor mass. 
First, when the progenitor mass increases, the number of small pulses
decreases while the number of big pulses increases. 
Second, the strength of the big pulses increase 
with the progenitor mass, which leads to a lower 
central temperature and density during its expansion. 
Third, the path during its early pulses becomes closer
to the PC instability as mass increases. Fourth, the 
second strong pulse strength is stronger than the first strong pulse. 

\subsection{Energetics}

\begin{figure*}
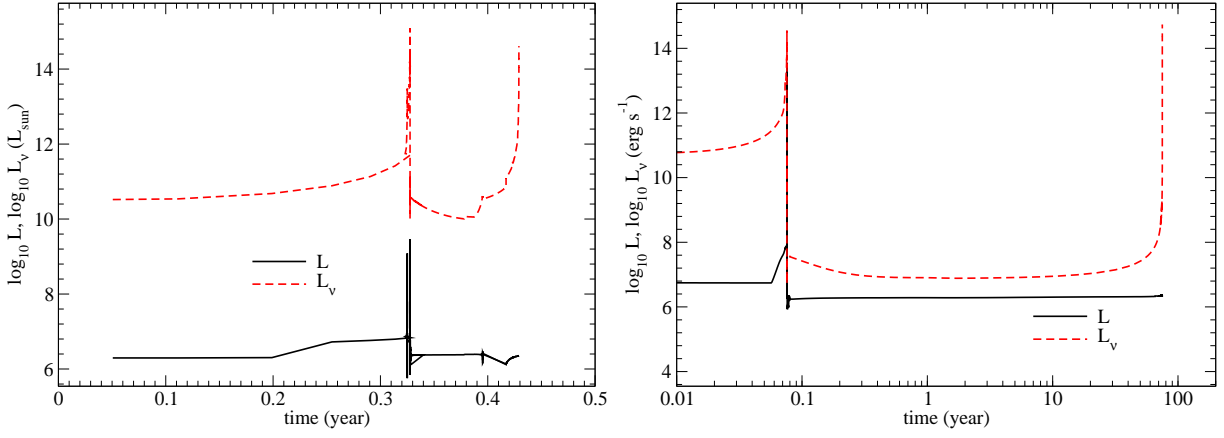

\centering
\includegraphics*[width=8cm,height=5.7cm]{fig21a.eps}
\includegraphics*[width=8cm,height=5.7cm]{fig21b.eps}

\caption{
Total luminosity and neutrino luminosity 
against time for Models He40A (left panel) and 
He45A (right panel) respectively.}
\label{fig:lumin_plot}
\end{figure*}

\begin{figure*}
\centering
\includegraphics*[width=8cm,height=5.7cm]{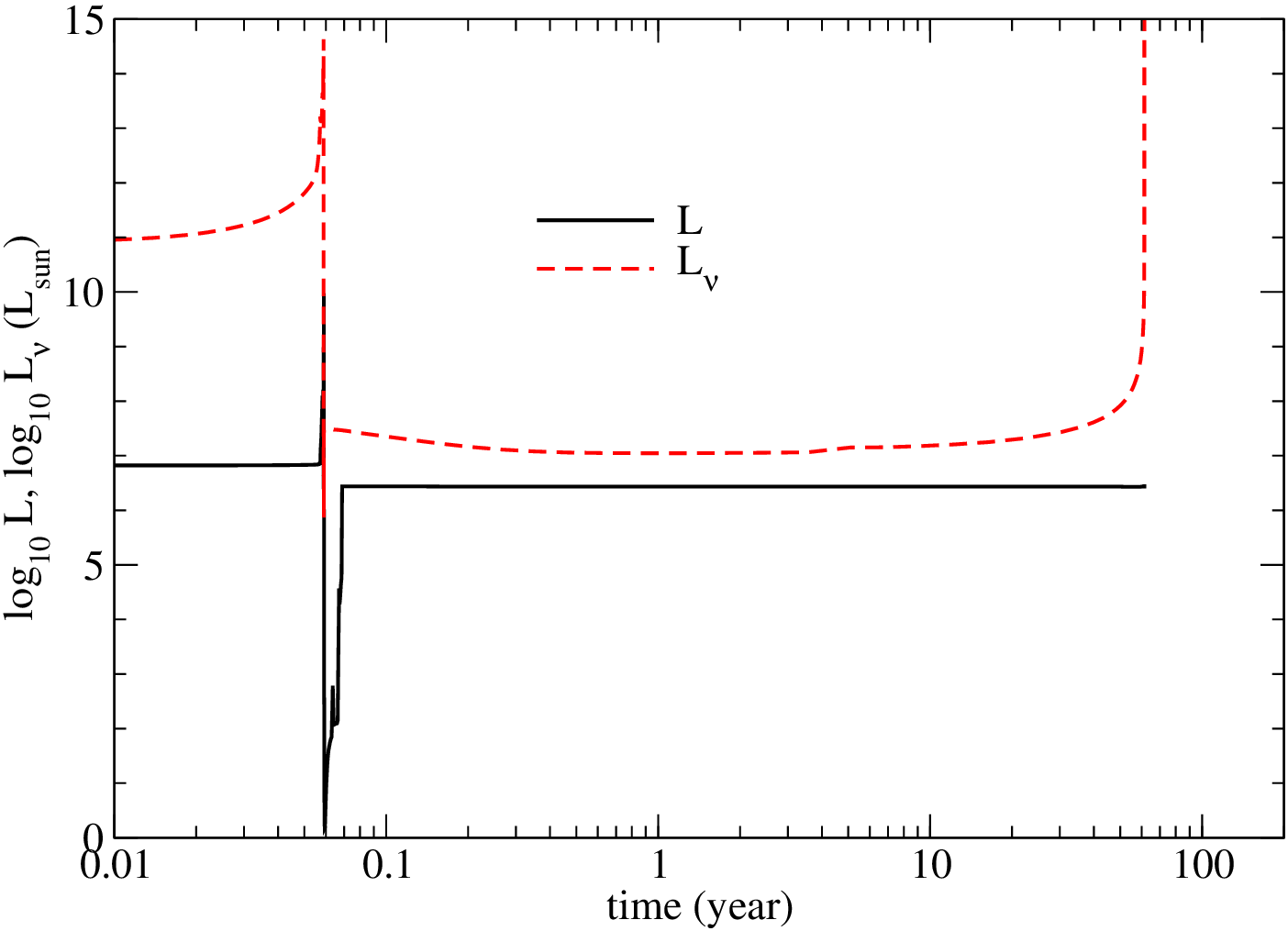}
\includegraphics*[width=8cm,height=5.7cm]{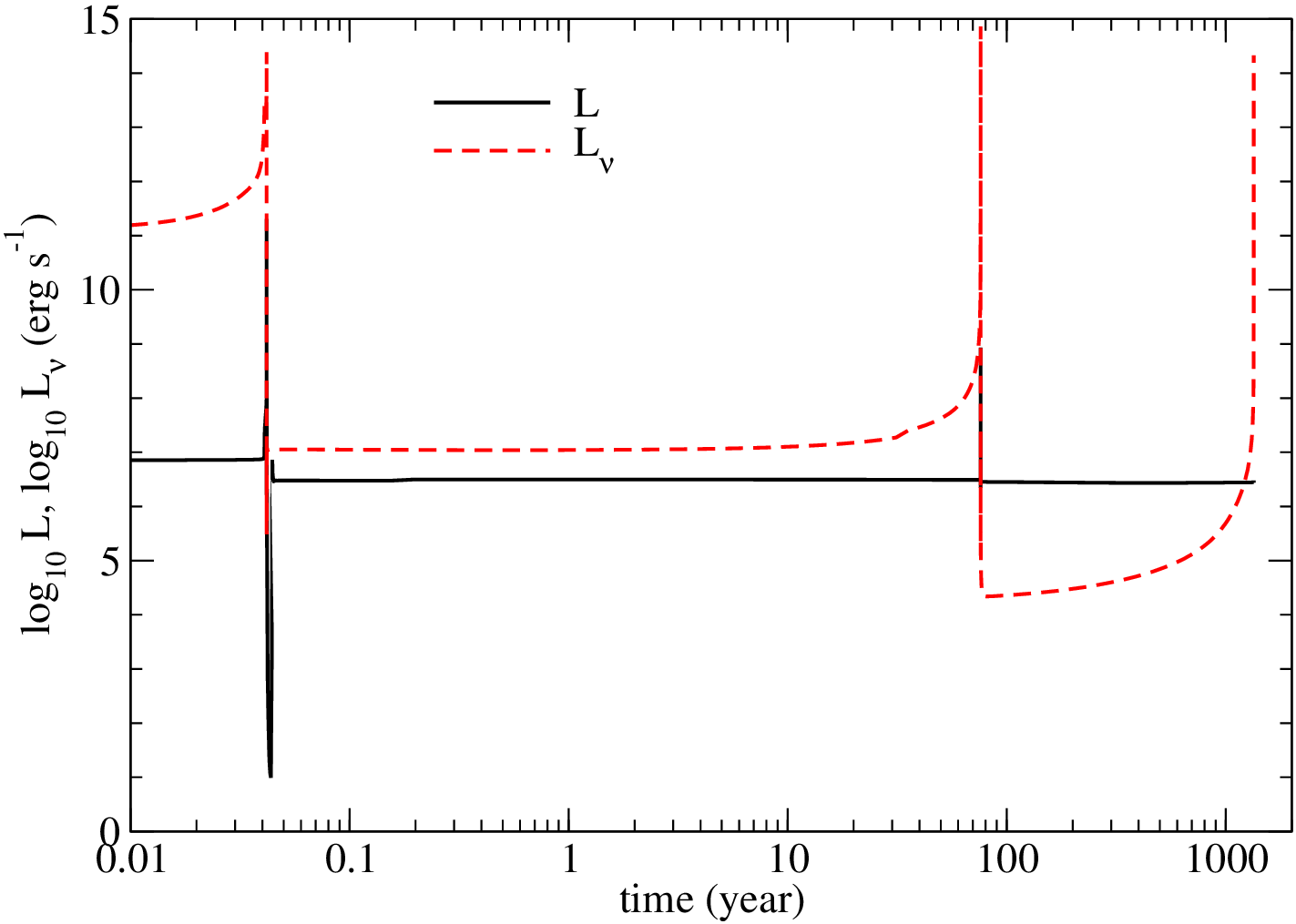}
\caption{
Similar to Figure \ref{fig:lumin_plot}, but for 
Models He50A (left panel) and
He55A (right panel) respectively.}
\label{fig:lumin_plot2}
\end{figure*}

\begin{figure*}
\centering
\includegraphics*[width=8cm,height=5.7cm]{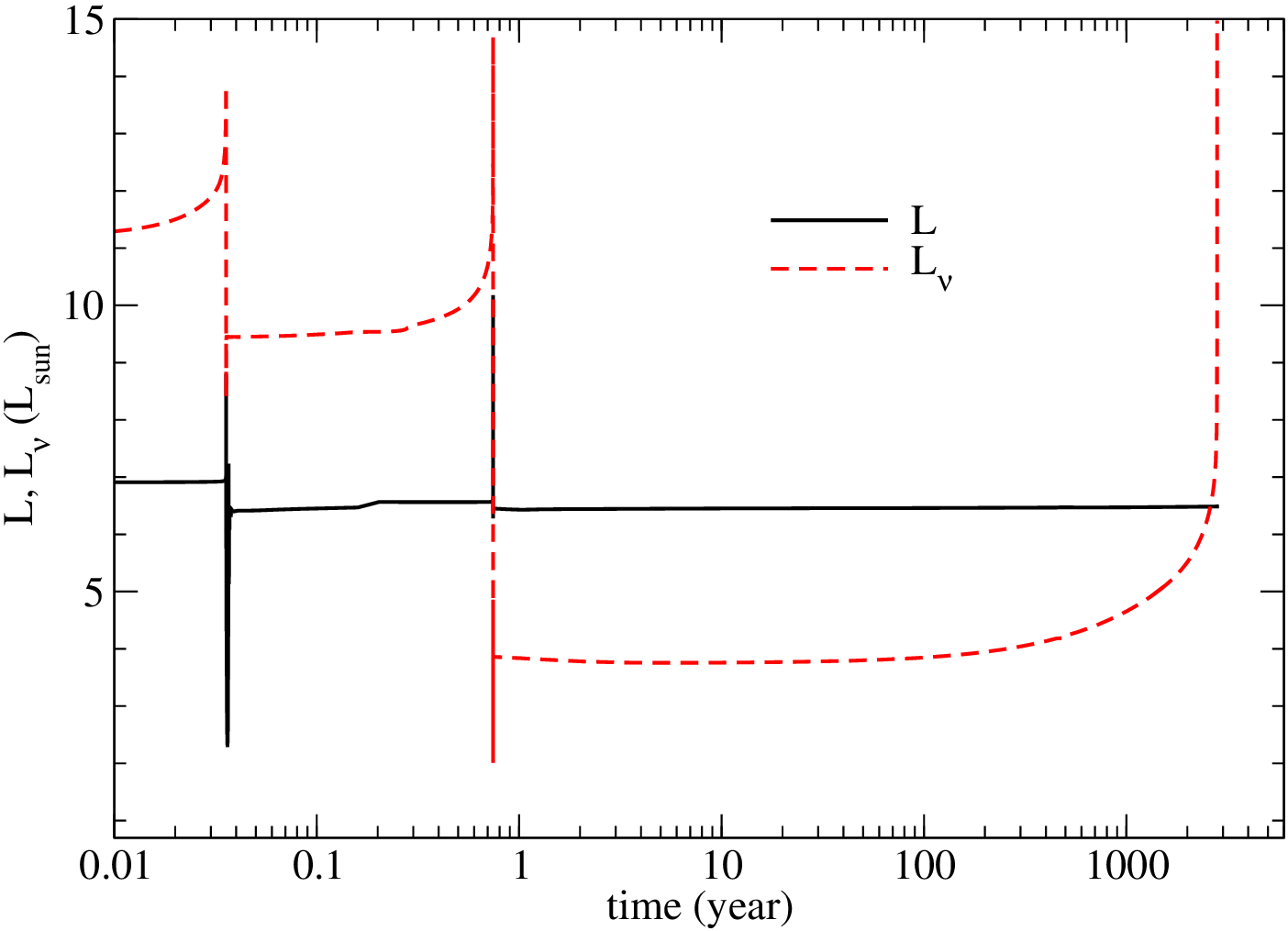}
\includegraphics*[width=8cm,height=5.7cm]{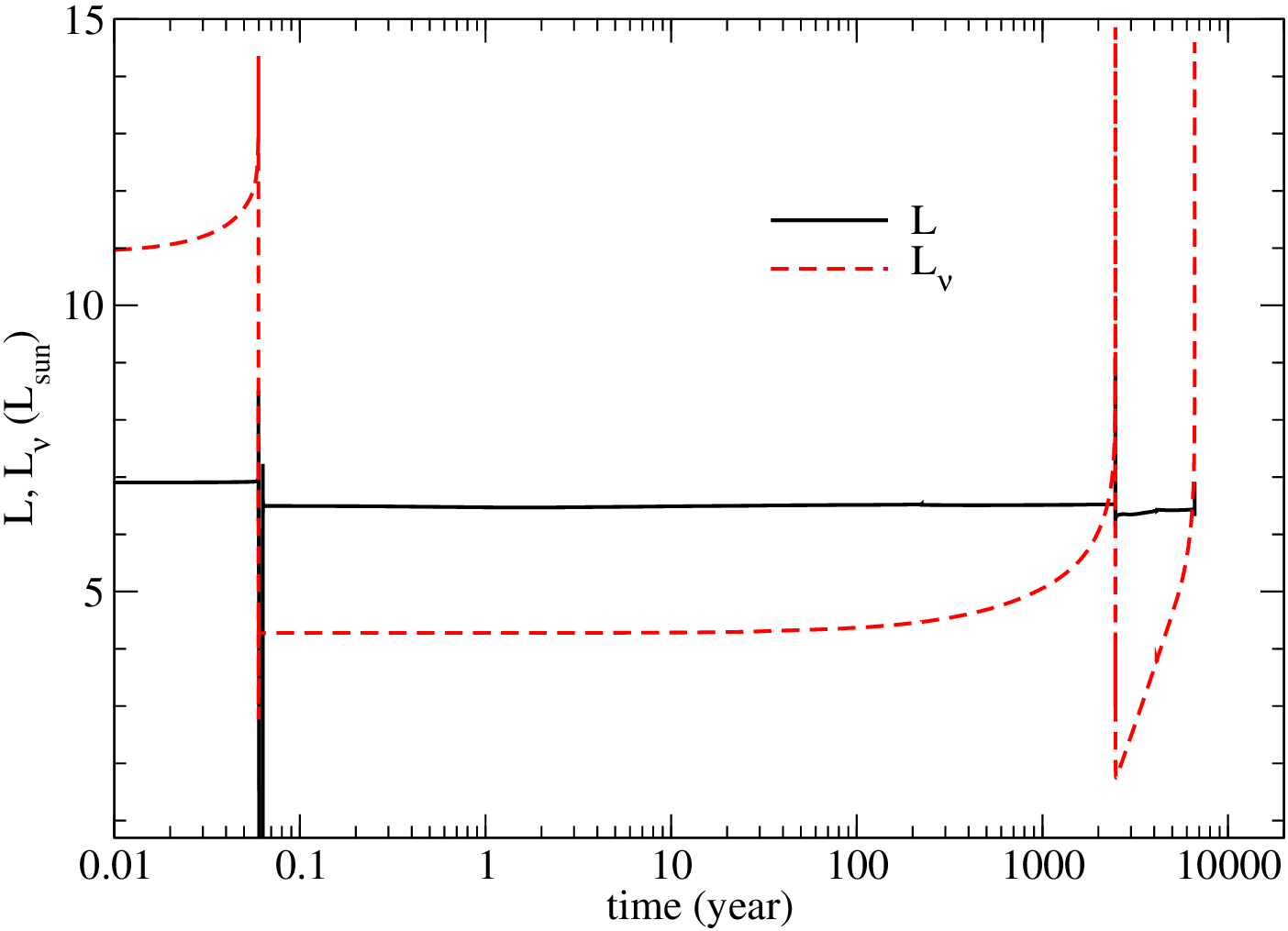}
\caption{
Similar to Figure \ref{fig:lumin_plot}, but for 
Models He60A (left panel) and He62A (right panel) respectively.}
\label{fig:lumin_plot3}
\end{figure*}

In Figures \ref{fig:energy_plot}, \ref{fig:energy_plot2}, and \ref{fig:energy_plot3} 
we plot the energy evolution for 
Models He40A, He45A, He50A, He55A, He60A and He62A, including
the total energy $E_{{\rm total}}$, 
internal energy $E_{{\rm int}}$, gravitational energy $E_{{\rm grav}}$ 
and kinetic energy $E_{{\rm kin}}$. The energy is scaled 
in order to make the comparison easier. 

In all six models, the energy evolution
does not depend on the stellar mass strongly, except for the 
energy scale. In all these models, 
the small pulses do not make observable changes
in the energy except for very small wiggles. 
The contraction before a pulse
leads to a denser and hotter core, where neutrino emission 
continuously draws energy from the system. At a big pulse, 
the total energy shows a rapid jump which increases
close to zero, then the ejection of mass quickly removes
the generated energy, making the star bounded again.
Similar jumps in $E_{{\rm int}}$ and $E_{{\rm grav}}$ show
that the core is strongly heated due to contraction 
heating and nuclear reactions. After that, the star reaches
a quiescent state with a mild increase of total energy
due to the $^{56}$Ni decay, then followed by a quick drop 
when it contracts again.

\subsection{Luminosity}

In Figures \ref{fig:lumin_plot}, \ref{fig:lumin_plot2} and \ref{fig:lumin_plot3}
we plot the luminosity evolution
for the six models similar to Figure \ref{fig:energy_plot}.
During the pulse, the extra energy from nuclear reactions
allows the luminosity to grow by 3-4 orders of magnitude. 
For a short period of time ($\leqslant 10^{-2}$ year), then 
the star becomes dim suddenly. After that
the star resumes its original luminosity quickly and
remains unchanged until the next pulse or final collapse. 
We remind that the luminosity during shock breakout
and shortly after it cannot be trusted because it requires
in general non-equilibrium radiative transfer for an accurate treatment.

The neutrino luminosity is more sensitive to the structure of
the star. The neutrino luminosity can also jump by 
3 -- 10 orders of magnitude from its typical luminosity in 
the hydrostatic phase to the maximally compressed state.
After the star has relaxed, the neutrino luminosity 
drops drastically. Depending on the strength of the pulse, 
neutrino cooling can become unimportant in the quiescent phase. 

\subsection{Mass Loss History}
\label{sec:massloss}

\begin{table*}
\begin{center}
\label{table:massloss_energy}
\caption{Energetic and chemical composition of the ejecta.
'Pulse' stands for the sequence of pulses in its evolution.
'Time' is the occurrence time in units of year. $T_{{\rm ej}}$
is temperature range of the ejecta in units of K. 
$E_{{\rm ej}}$ is the ejecta energy in units of $10^{50}$ erg.
$M$(He), $M$(C), $M$(O), $M$(Ne), $M$(Mg), $M$(Si) are the 
masses of He, C, O, Ne, Mg and Si in the ejecta in units of 
solar mass.}
\begin{tabular}{|c|c|c|c|c|c|c|c|c|c|c|c|}
\hline
Model & Pulse & time & $M_{{\rm ej}}$ & $E_{{\rm ej}}$ & $T_{{\rm ej}}$ & 
$M$(He) & $M$(C) & $M$(O) & $M$(Ne) & $M$(Mg) & $M$(Si) \\ \hline
He40A & 6 & $9.9 \times 10^{-2}$ &  1.0 &  1.0 & 6.3-6.8 & 1.0 & 0.0 &  0.0 & 0.0 & 0.0 & 0.0 \\ \hline
He45A & 1 & $7.4 \times 10^{-2}$ &  4.0 &  6.6 & 6.5-6.9 & 3.8 & 0.2 &  0.0 & 0.0 & 0.0 & 0.0 \\ \hline
He50A & 2 & $2.0 \times 10^{-1}$ &  4.0 &  2.5 & 6.7-7.2 & 3.9 & 0.1 &  0.0 & 0.0 & 0.0 & 0.0 \\ \hline
He55A & 1 & $6.2 \times 10^{-2}$ &  0.3 &  1.8 & 6.8-7.1 & 0.3 & 0.0 &  0.0 & 0.0 & 0.0 & 0.0 \\
He55A & 2 & $1.9$                & 10.0 & 13.1 & 6.0-6.7 & 7.5 & 1.0 &  0.8 & 0.2 & 0.3 & 0.2 \\ \hline
He60A & 1 & $1.7 \times 10^{-1}$ & 10.6 &  5.1 & 5.3-6.4 & 8.6 & 2.4 &  0.8 & 0.6 & 0.2 & 0.2 \\
He60A & 2 & $7.4 \times 10^3$    & 38.7 & 59.0 & 6.0-6.5 & 0.0 & 1.1 & 32.5 & 1.3 & 1.3 & 2.0 \\ \hline
He62A & 1 & $5.5 \times 10^{-2}$ &  0.6 &  0.1 & 6.8-7.4 & 0.6 & 0.0 &  0.0 & 0.0 & 0.0 & 0.0 \\
He62A & 1 & $3.5 \times 10^{ 3}$ & 55.4 & 29.6 & 4.6-7.2 & 7.8 & 1.7 & 33.2 & 1.6 & 1.8 & 5.8 \\ \hline
He64A & 1 & $4.9 \times 10^{-2}$ & 21.8 & 29.4 & 4.4-7.1 & 8.5 & 1.8 &  9.9 & 0.9 & 0.3 & 0.6 \\ \hline

\end{tabular}
\end{center}
\end{table*}

\begin{figure*}
\centering
\includegraphics*[width=15cm,height=10cm]{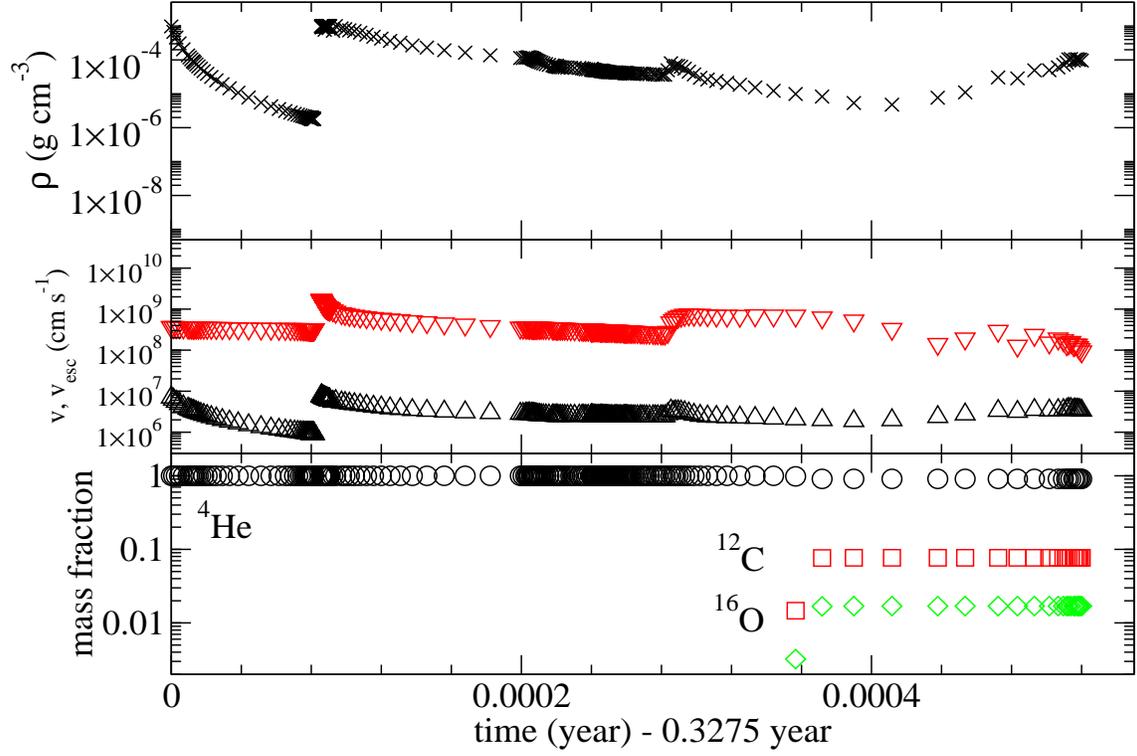}

\caption{
The mass ejection history of Model He40A, including 
in the upper panel the ejecta density, in the middle panel the 
ejecta velocity and the escape velocity and in the lower panel
the ejecta chemical composition.
}
\label{fig:ejecta_plot}
\end{figure*}

\begin{figure*}
\centering
\includegraphics*[width=15cm,height=10cm]{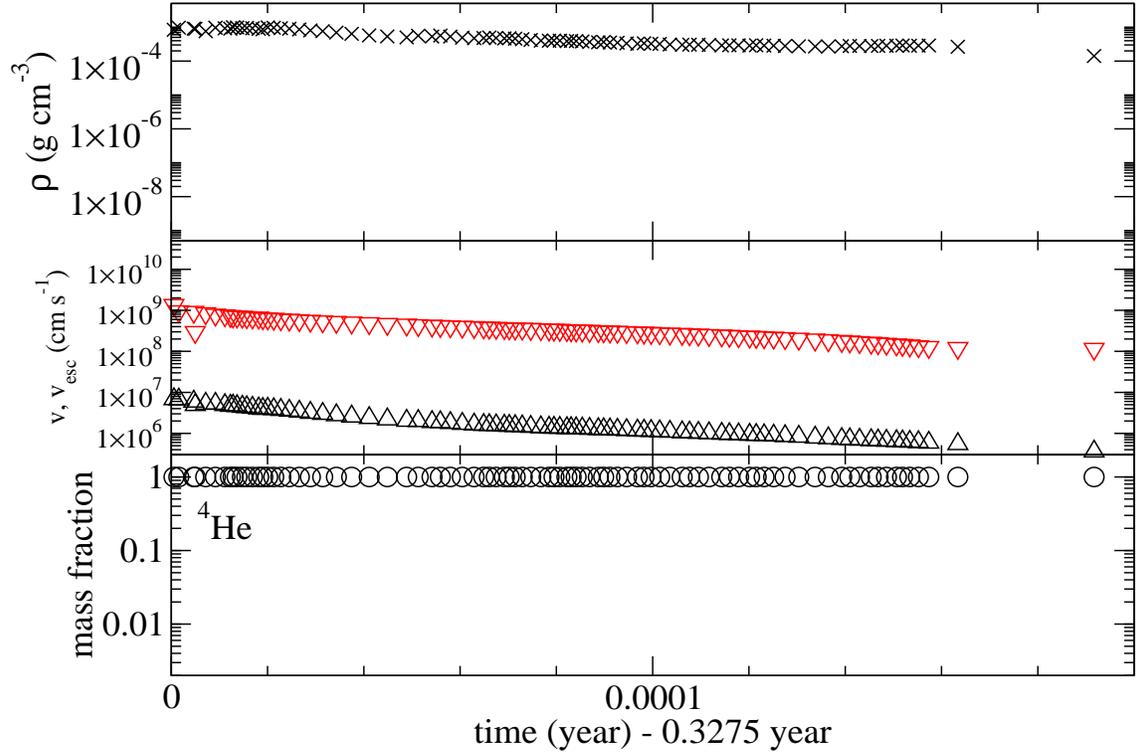}
\caption{$(cont'd)$
The mass ejection history of Model He50A including the 
density (top panel), velocity (middle panel) and chemical composition (bottom panel).
For the velocity plot, the black triangles and red inverted triangles
correspond to the ejecta and escape velocities at the surface.
}
\label{fig:ejecta_plot2}
\end{figure*}

\begin{figure*}
\centering
\includegraphics*[width=15cm,height=10cm]{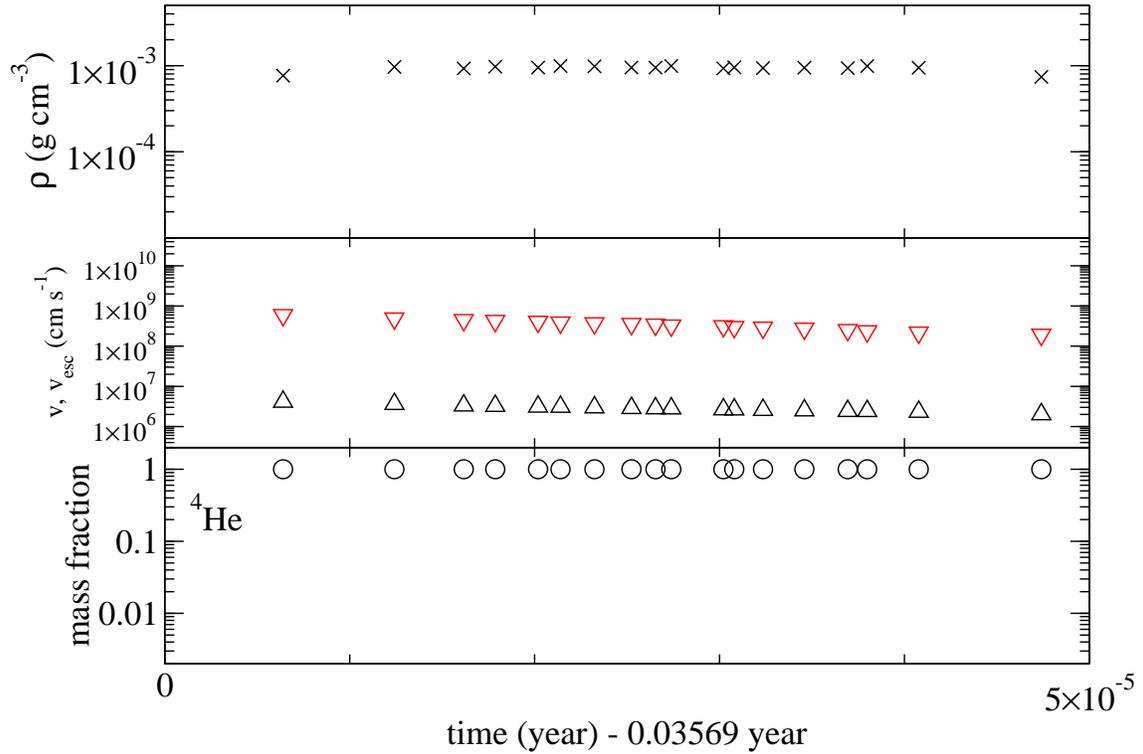}
\caption{$(cont'd)$
The mass ejection history of Model He60A including the 
density (top panel), velocity (middle panel) and chemical composition (bottom panel).
}
\label{fig:ejecta_plot3}
\end{figure*}

\begin{figure*}
\centering
\includegraphics*[width=15cm,height=10cm]{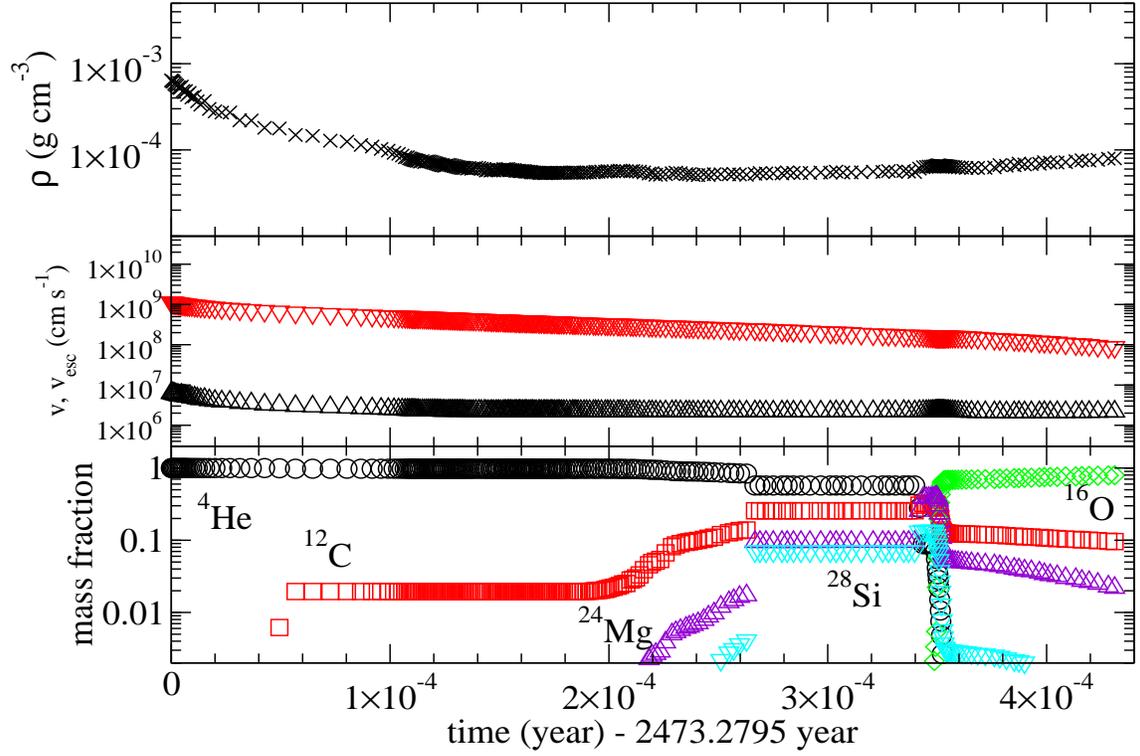}

\caption{$(cont'd)$
The mass loss history of the second strong pulse in Model He62A.}
\label{fig:ejecta_plot4}
\end{figure*}

\begin{figure*}
\centering
\includegraphics*[width=15cm,height=10cm]{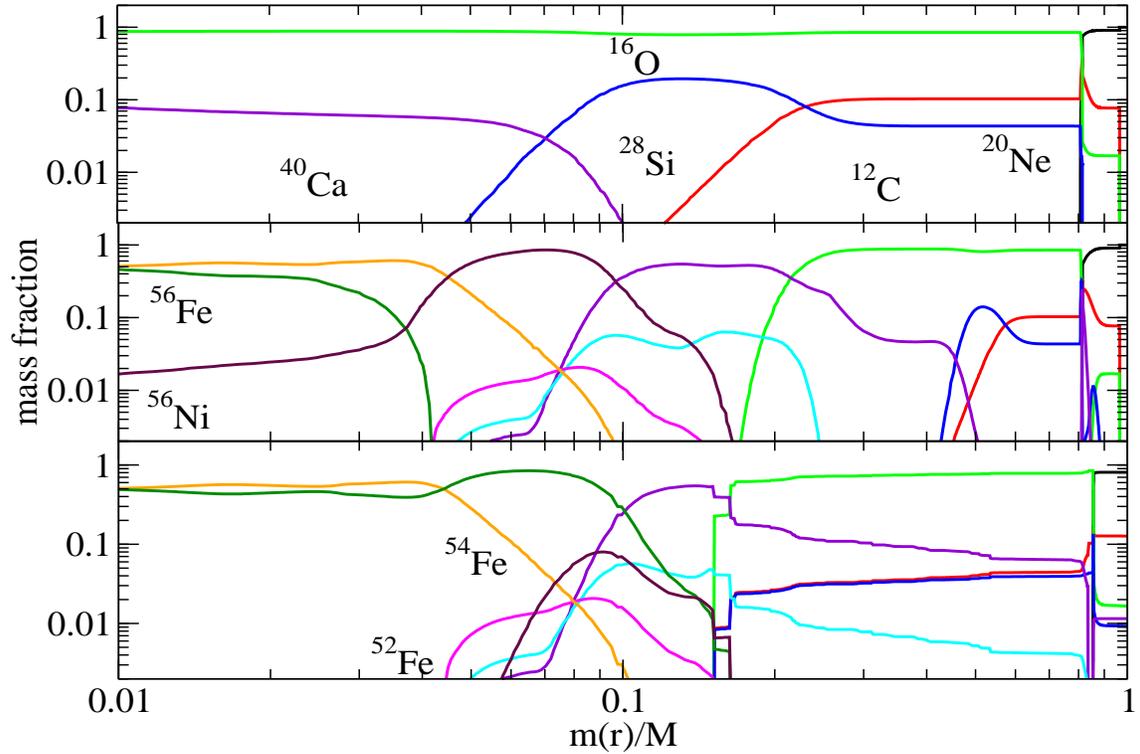}

\caption{
The chemical composition of Model He40A before the
first pulse, after the first pulse and before the final pulse 
in the upper middle and lower plot. Here we define the 
star entering the pulsation phase when the core reaches
$10^{9.3}$ K. 
}
\label{fig:xiso_plot}
\end{figure*}

\begin{figure*}
\centering
\includegraphics*[width=15cm,height=10cm]{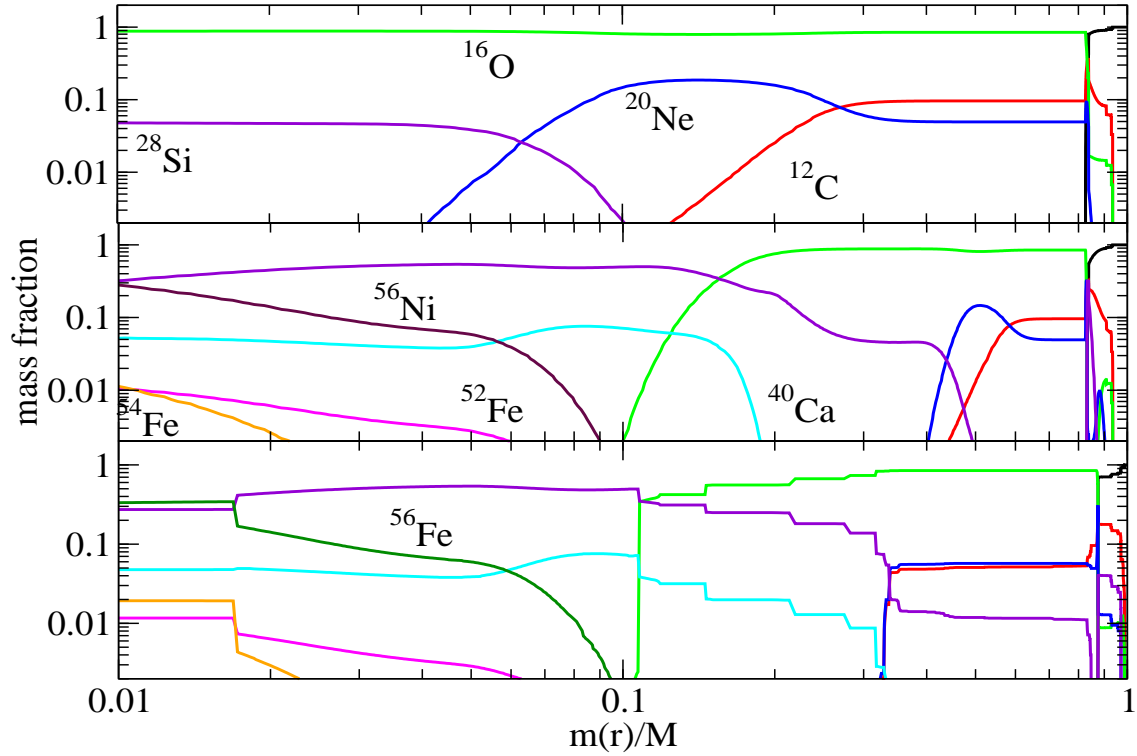}
\caption{$(cont'd)$
Similar to Figure \ref{fig:xiso_plot}, but for Model He50A
before the first pulse, after the first pulse and before the final pulse 
in the upper middle and lower plot.
}
\label{fig:xiso_plot2}
\end{figure*}

\begin{figure*}
\centering
\includegraphics*[width=15cm,height=10cm]{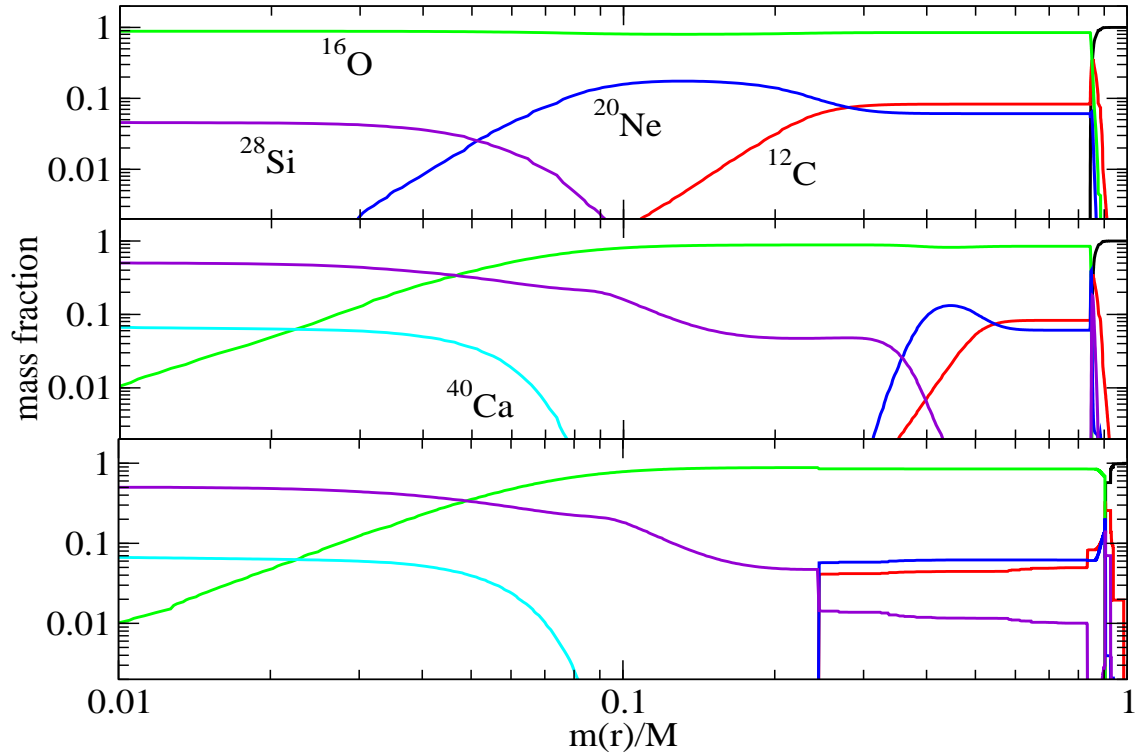}

\caption{$(cont'd)$
The abundance patterns for Model He62A
before the first pulse, after the first pulse and before the second pulse.}
\label{fig:xiso_plot3}
\end{figure*}

\begin{figure*}
\centering
\includegraphics*[width=15cm,height=10cm]{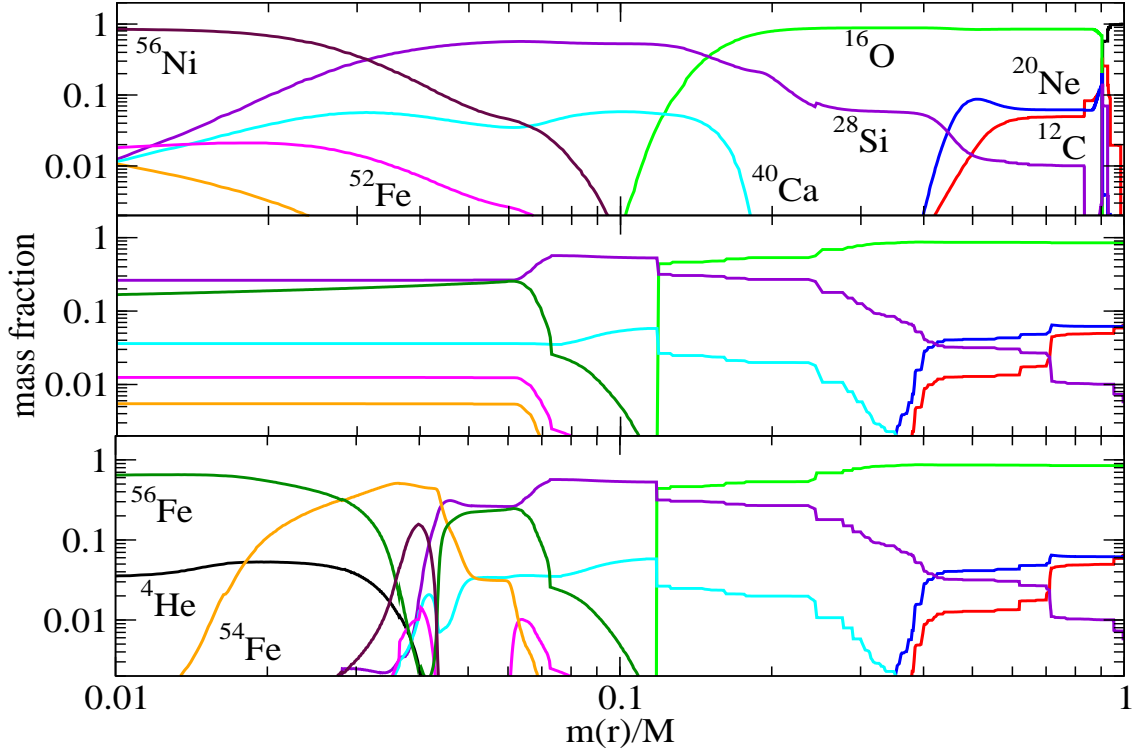}

\caption{$(cont'd)$
(Similar to Figure \ref{fig:xiso_plot3}, but for Model He62A
after the second pulse, before the final pulse and at the end of simulation.}
\label{fig:xiso_plot4}
\end{figure*}

During the pulsation, when the bounce leads to the explosive
burning in the core and inner envelope,
sufficient energy is produced to create an outgoing shock,
where the outermost matter can gain sufficient energy to be ejected from the star. The ejected matter 
later cools down and becomes the circumstellar matter (CSM). 
The existence of such H-free CSM is necessary in the circumstellar interaction models for
Type I superluminous supernovae (SLSNe-I) \citep{Sorokina2016, Tolstov2017}.
The chemical and hydrodynamics properties of the CSM thus
become important, which influence the formation of the light curve
of the explosion. 

In Table \ref{table:massloss_energy} 
we tabulate the mass loss history of each model and its chemical composition. 
In Figures \ref{fig:ejecta_plot}, \ref{fig:ejecta_plot2}, \ref{fig:ejecta_plot3} 
and \ref{fig:ejecta_plot4} we plot the ejecta profiles
of the representative pulsation taken from
Models He40A, He50A He60A and He62A
respectively. Three patterns can be observed in 
the mass ejection. We choose these models because these
examples characterize the typical ejecta features
of strong pulses in the lower 
and higher mass regimes.
We take the numerical values when the mass shells
are ejected during the pulsation because that is the last moment
the code keeps track of their evolution.

The first group is the strong pulse in the 
lower mass branch. 
In Models He40A and He45A the last pulse is 
the pulse which ejects mass.
It shows wiggles in its density profiles, showing that
the thermal expansion creates the first wave of mass
ejection, followed by the shock as the velocity 
discontinuity approaching the surface, which creates the 
second wave of mass ejection. In both cases, only 
the He layer is affected, but as the He layer becomes
thin, matter near the CO layer is ejected. 

The second group is the weaker pulse 
of the more massive branch.
In Models He50A, He55A He60A and He62A 
the first strong pulse occurs after the core 
starts to consume O collectively. Since it burns 
much less O than other strong pulses, 
the ejection comes from the rapid expansion of the star,
which includes matter in the He envelope. 

The third group is the strong pulse of the 
more massive branch. 
In Models He55A and Model He60A, the second pulse
is stronger so that the ejecta density 
gradually decreases. A continuous ejection
of mass in terms of smooth density profile
is found. The mass ejection is sufficient 
deep that at the end of pulsation, traces of 
$^{12}$C and $^{16}$O can be found. We remark
that the inclusion of massive elements (compared
with H and He) will be important for the 
future light curve modelling because they contribute
as the main source of opacity. 

One of the pulsations needs to be discussed separately
because of its very massive mass ejection, which 
involves very unique chemical composition in its ejecta. 
In Model He62A (right plot), the second pulse
becomes strong enough that, besides its decreasing
density profile, the later ejected material contains 
a significant amount of heavier elements including
C, O, Mg and Si, showing that the He envelope is 
completely exhausted before the star is sufficiently relaxed.

\subsection{Chemical Properties}

In Figures \ref{fig:xiso_plot}, \ref{fig:xiso_plot2}, \ref{fig:xiso_plot3}
and \ref{fig:xiso_plot4} we plot the isotope profiles
at different moments of selected Models He40A, He50A, and
He62A. We selected moments before and after
each strong pulse to extract the nuclear burning history. 
The models are chosen to demonstrate how different strength of 
pulsation and its convective mixing between pulses
can create distinctive isotope abundances in the star. 
By comparing the isotope distribution, we can understand 
which part of burning contributes to the evolution of pulsation. 

In all models, it can be seen that the star is simply a
pure O-core with a minute amount of Si in the core
or C in the envelope, covered by a pure He surface. 
However, their changes can be very different depending 
on the progenitor mass.

In Model He40A, after the strong pulse, due to its previous
weak pulses which continue to burn matter in the core, 
a range of elements are produced including $^{52}$Fe, $^{54}$Fe, 
$^{56}$Fe and also $^{56}$Ni. There is a clear structure for 
each layer, which comes in the order of $^{56}$Fe, 
$^{54}$Fe, $^{56}$Ni, $^{40}$Ca, $^{16}$O and then $^{4}$He.
After that, the core relaxes and becomes quiescent until
it completely loses its thermal energy produced during
the pulse, while at the same time convection re-distributes
the matter for a flat composition profile. 
Most convection occurs at $q > 0.1$, where the
convective shells of different sizes make a staircase like
structure. 

Models He45A and He50A share a similar nuclear 
reaction pattern. We choose Model He50A as an example.
The strong pulse provides the 
required temperature and density to make Ni 
in the center and Si in the outer zone. The Si-rich 
zone extends to $q \approx 0.2$. During the quiescent
phase, convection not only mixes the material 
in the envelope, but also in the core, which is
seen by the stepwise distribution of $^{52}$Fe 
and $^{54}$Fe. 

In Models He55A, He60A and He62A share also similar 
abundance pattern. We choose the evolution of He62A
as an example. The first pulse makes
the original O-core into mostly Si and some Ca. 
Again, the convective mixing during the quiescent state 
redistributes the matter near the surface $(q > 0.25)$. 
In the second strong pulse, the nuclear reaction is
very similar to the late pulses of Model He40A
and He45A. Ni forms in the innermost part, with
a small amount of Fe isotopes like $^{52}$Fe and $^{54}$Fe. 
Then Si and Ca form the middle layer and at last
the He envelope appears. During the quiescent phase, the
convection occurs in a deeper layer compared
which is absent in lower mass models.

\section{Models for Super-Luminous Supernovae}
\label{sec:SLSN}

PPISNe have been used to model the 
superluminous supernovae
such as SN2006gy \citep{Woosley2007},
SN2010gx, PTF09cnd \citep{Sorokina2016} and PTF12dam \citep{Tolstov2017}. 
PTF09cnd and PTF12dam are challenging as
they require such massive CSM as 
20 -- 40 $M_{\odot}$ CSM prior to the 
supernova explosions. Furthermore, these SLSNe are of Type I so that CSM needs 
to be H-free with the presence of He, C and O in order
to explain the high opacity surrounding.

Models with $\sim$ 64 $M_{\odot}$ He core is likely to 
eject a mass $\sim 22 ~M_{\odot}$. 
Our model gives an ejecta with He, C and O masses 
of 8.5, 1.8, 9.9 $M_{\odot}$. The corresponding ratio
of He:C is therefore 4.8:1:5.5. This is close to the values 
in the model M66R170E27(CSM19) of \cite{Tolstov2017}, which has an abundance of C:O = 1:4.
Whether the following collapse of this
$\sim 40 ~M_{\odot}$ remnant can explode energetically
with an energy of $2 - 3 \times 10^{52}$ erg is uncertain. 

\begin{figure}
\centering
\includegraphics*[width=7cm,height=5cm]{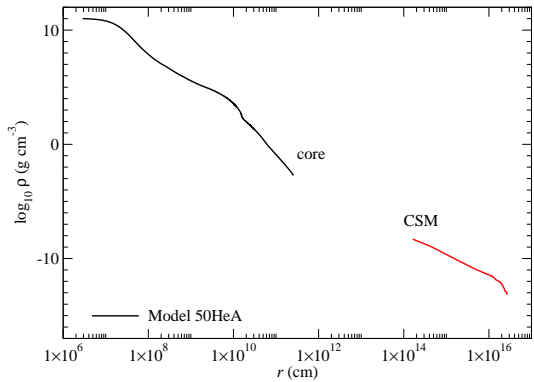}
\caption{The final density profile for Model He50A including
the core and ejecta matter (CSM) at the onset of collapse.}
\label{fig:CSM_plot2}
\end{figure}

In Figure \ref{fig:CSM_plot2} we plot the density profile
of Model He50A at the onset of collapse ($\rho_c = 10^{10}$ g cm$^{-3}$).
Both the CSM and core are included. The CSM is constructed 
from the mass ejection history, which is obtained by
a post-process manner until the core begins to collapse.
The core consists of 
a compact Fe core of $\sim 2 ~M_{\odot}$ with $r < 10^8$ cm.
Outside the Fe core, a smooth Si-rich envelope extends up to 
$M(r) \sim ~10 M_{\odot}$ and $r \sim 10^{10}$ cm.
The outer surface extends to $\sim 10^{12}$ cm.
We remark that the outermost envelope is mostly the remaining
matter which is not ejected near the end of pulsation
event. They are mostly decoupled gravitationally from the 
core. The original stellar envelope is the middle envelope
of the final density profile. 

As discussed in Section \ref{sec:massloss}, the mass ejection
of He50A is smooth which occurs for 0.0002 year ($\sim 2$ hours) 
before the outermost shell is bounded. Such continuous mass
ejection can produce a smooth and extended CSM outside the 
star. There is no significant
collision among ejected masses, where the collision can give rise to observable density discontinuities. 
Without mass collision, 
the CSM profile in general follows the $1/r^2$ scaling,
which extends from $10^{14}$ to $10^{16}$ cm. We note that in the calculation,
there is a gap between the outer envelope of the core to the inner
boundary of the CSM from $10^{11}$ to $10^{14}$ cm.
There should be fallback by gravitational tidal force
on the ejecta. However, to resolve this one requires 
another hydrodynamical experiment to follow how the 
ejecta exchanges momentum.

\section{Black Hole Masses from Pulsational Pair-Instability Supernovae}
\label{sec:BH}

The gravitational wave detectors aLIGO and VIRGO have recently
detected gravitational wave signals from 
merger events of compact objects. Some massive
black holes, for example in GW 150914, the 
black hole masses of $35.6 \pm ^{4.6}_{3.0}$ and $30.6 \pm^{3.0}_{4.4} ~M_{\odot}$
are measured \citep{Abbott2016b}. Another massive black hole
merger event is GW 170104, where the binary consists of
black holes of masses 31.0 and 20.1 $M_{\odot}$ respectively.
In Table \ref{table:BH_mass} we list out the recent 
gravitational wave events with black hole masses reaching
above $30 ~M_{\odot}$ within one sigma. 
A recent statistics has further pushed the maximum
pre-merger black hole mass to $\sim$ 55 $M_{\odot}$ \citep{Abbott2018}.
It is unclear, whether the massive black hole
forms directly from the collapse of a massive star, or 
has experienced multiple merger events prior to the
event detected by the gravitational wave detectors.

\begin{table}
\label{table:BH_mass}
\begin{center}
\caption{The primary and secondary black hole masses with one
or both black hole masses exceeding $30 ~M_{\odot}$ within 
one sigma. Events in bold font are those with black hole masses
exceeding $40 ~M_{\odot}$.
$m_1$ and $m_2$ are the black hole masses
in units of $M_{\odot}$. The data are taken from \cite{Abbott2018}.}

\begin{tabular}{|c|c|c|}
\hline
Event & $m_1$ & $m_2$ \\ \hline
\textbf{GW150914} & $35.6 \pm ^{4.8}_{3.0}$ & $30.6 \pm ^{3.0}_{4.4}$ \\
GW151012 & $23.3 \pm ^{14.0}_{5.5}$ & $13.6 \pm ^{4.1}_{4.8}$ \\
GW170104 & $31.0 \pm ^{7.2}_{5.6}$ & $20.1 \pm ^{4.9}_{4.5}$ \\
\textbf{GW170729} & $50.6 \pm ^{16.6}_{10.2}$ & $34.3 \pm ^{9.1}_{10.1}$ \\
\textbf{GW170809} & $35.2 \pm ^{8.3}_{6.0}$ & $23.8 \pm ^{5.2}_{5.1}$ \\
GW170814 & $30.7 \pm ^{5.7}_{3.0}$ & $25.3 \pm ^{2.9}_{4.1}$ \\
\textbf{GW170818} & $35.5 \pm ^{7.5}_{4.7}$ & $26.8 \pm ^{4.3}_{5.2}$ \\
\textbf{GW170823} & $39.6 \pm ^{10.0}_{6.6}$ & $29.4 \pm ^{6.3}_{7.1}$ \\ \hline

\end{tabular}
\end{center}
\end{table}

Our model suggests that
the single star scenario has an upper limit for the black hole mass. 
He core with a mass greater than 64 $M_{\odot}$, 
the star does not collapse, but 
explode as a pair-instability supernova. The collapse
only reappears for a star with a mass larger than 260 $M_{\odot}$
(for zero metallicity) \citep{Heger2002}. The corresponding
black hole mass is $\sim 100$ $M_{\odot}$.

\begin{figure*}
\centering
\includegraphics*[width=10cm,height=7cm]{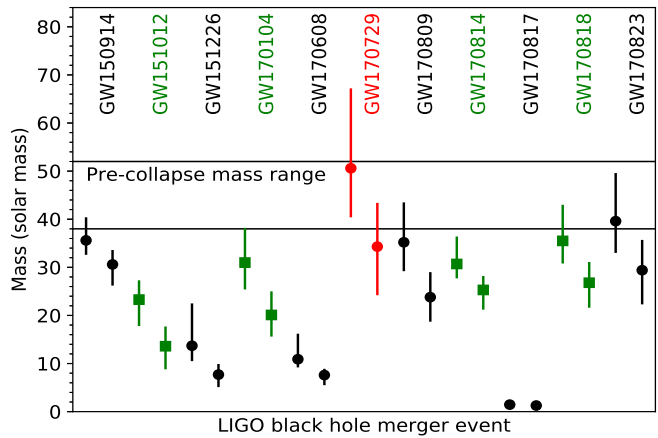}
\caption{The pre-collapse mass of the PPISN against progenitor
mass with the measured black hole masses obtained from
binary black hole merger events \citep{Abbott2016a,Abbott2016b,Abbott2017,Abbott2018}.
The left and right data points correspond to the primary and secondary 
black holes respectively. The error bars for the pre-merge neutron star 
for GW170817 are too small
to be seen in the current scale.}
\label{fig:BHmass_M_plot}
\end{figure*}

To connect PPISN with the measured black hole mass spectra, 
we plot in Fig. \ref{fig:BHmass_M_plot} the remnant mass 
against progenitor mass, and the mass range of the black hole
measured by the gravitational wave signals. 
He cores with a mass between 40 -- 64 $M_{\odot}$, a
mass correction is included to account for the 
pulsation-induced mass loss. Beyond $M_{{\rm He}} = 64 ~M_{\odot}$
the star enters the pair-instability regime and no 
compact remnant object is left. Near $\sim 62$ $M_{\odot}$
the remnant mass is the maximum at $\sim 52 ~M_{\odot}$. 
Some of the events can be explained by the current 
PPISN picture. This includes the primary black hole in the events
GW150914, GW170104, GW170729, GW170809, GW170818 and GW170823
and the secondary black hole in the event GW170729. 

We remark that there exists high uncertainties in how to 
connect the He core mass with the final remnant mass. 
In our simulations, the pulsation induced mass loss
is done in one-dimension. When the multi-dimensional
effects, e.g. Rayleigh-Taylor instabilities, can be considered 
during the propagation of pulse, the actual mass loss can be changed.
Also, after the Fe core collapses, during formation of proto-neutron
star and black hole,
the mass ejection and neutrino energy
may reduce the final remnant mass by $\sim 10 \%$ (See e.g. \cite{Zhang2008,Chan2018}). 
We remark that interpreting the He core mass
as the final remnant mass can only be an upper limit
of the black hole mass. 
The black hole accretion disk around a rotating black hole
allows formation of high velocity jet.
The magnetohydrodynamical instability of the accretion disk 
can easily fragment the disk and send the energetic jet
to the stellar envelope. This process can lower the remnant
mass. Therefore, as a first 
estimation of our result, we use the He core mass
as an upper estimate of the final black hole mass.

We note that in a single observation, the solution for matching
the black hole mass with our remnant mass is degenerate
for both mass and metallicity. To further apply the 
black hole information in PPISN to constrain the mass loss,
population of black hole mass will become important, which 
can directly constrain the current mass loss model, when 
combined with suitable stellar initial mass functions. 

\begin{figure*}
\centering
\includegraphics*[width=10cm,height=7cm]{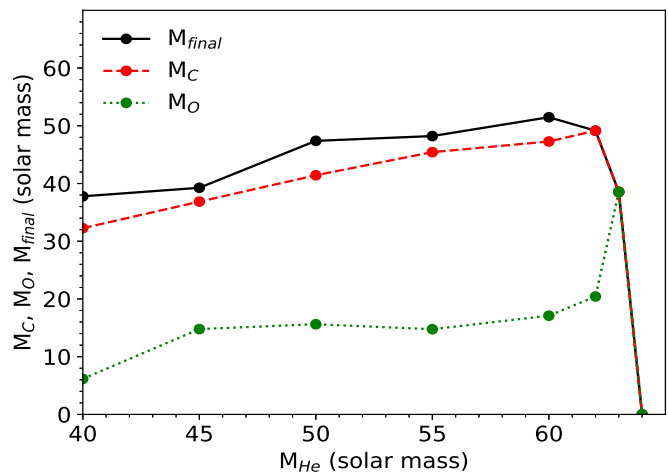}
\caption{The pre-collapse mass of the PPISN, C- and O- core mass against progenitor
He-core mass. Here we define the C- and O-cores to be the inner boundaries
where the local $^{4}$He and $^{12}$C mass fractions drop below 
$10^{-2}$.}
\label{fig:BHmass_M_plot2}
\end{figure*}

In Figure \ref{fig:BHmass_M_plot2} we plot the 
final stellar mass, C- and O-core masses
for all the He-core models. We define the 
boundary of the C- and O-cores to be the 
inner boundaries where the local $^{4}$He and $^{12}$C mass fractions drop below 
$10^{-2}$. We can see that three
layers appear. For $M_{{\rm He}} = 40$, 45 and 
50 $M_{\odot}$, there are explicit He-envelope, C- and
O-layers. For $M_{{\rm He}} = 55$ and 60 $M_{\odot}$, the huge
mass loss completely ejects the pure He layer,
which exposes the C-rich layer (combined with He). 
At $M_{{\rm He}} = 63$ $M_{\odot}$, the mass ejection
further shreds off the C-rich layer, exposing 
the O-layer. The whole star has everywhere 
the mass fraction of $^{12}$C below $10^{-2}$. 
Therefore, the He-core and C-core 
masses coincide with the stellar total mass. 
From this we can see to what level the mass
ejection takes place for the PPISN models. 
However, the definition of 
He- and C-core masses can be ambiguous at the end
of simulations because the matter becomes O-rich before 
C is exhausted. Similarly on the surface there can be 
non-zero abundance of $^{12}$C instead of pure $^{4}$He.  

\section{Conclusions}
\label{sec:conclusion}

In this paper, we studied pulsational
pair-instability (PPI) which occur in the He-core 
of $M_{\rm He} = 40 - 64 M_{\odot}$. These are the cores of 
80 -- 140 $M_{\odot}$ main-sequence stars. 
We used the one-dimensional 
stellar evolution code MESA and applied the 
implicit hydrodynamics module implemented
in the version 8118. 

\noindent (1) First, we computed the evolution of stars 
with the initial masses of 80 -- 140 $M_{\odot}$ and 
metallicities of $Z = 0.01 - 1 ~Z_{\odot}$
from the pre-main-sequence until the 
central temperature reaches $10^{9.3}$ K.
We examined how the final He- and CO-core mass
depends on the metallicity.  The star with a 
higher metallicity has a stronger stellar
wind mass loss, thus forming a smaller mass He-core.
In order for the star to form a He-core of more massive than
$40 ~M_{\odot}$ and thus to undergo PPI, $Z \lesssim 0.5 ~Z_{\odot}$ 
is required.

\noindent (2) We calculated the evolution of the He-cores 
of $M_{\rm He} = 40 - 64 ~M_{\odot}$ with $Z = 0$ from the He main-sequence 
through the onset of collapse.
These He cores undergo PPI.
We calculated  the hydrodynamical evolution of PPI 
with mass ejection.  
We examined nucleosynthesis during PPI, showing how each pulsation
changes the chemical composition of the star
and how the later convection alters the post-pulsation
star.

\noindent (3) The total ejected mass is almost a monotonically 
increasing function of $M_{\rm He}$ except for some 
fluctuations in the lower mass end. 
The He-core with a higher mass has fewer weak pulses that
do not eject masses. Instead it has much stronger pulses 
eject masses. 
The number of pulses ranges from 
6 weak pulses for $M_{\rm He} = 40 ~M_{\odot}$
to no weak pulse but 2 strong pulses for $M_{\rm He} = 62 ~M_{\odot}$.
The ejecta mass is lower than $1 ~M_{\odot}$
in the low mass end and increases to as large as $\sim 10 ~M_{\odot}$
near the pair-instability supernova regime.
Models with $M_{\rm He} > 64 ~M_{\odot}$ behave as pair-instability
supernovae, where no remnant is left.

\noindent (4) The ejecta form circumstellar
matter (CSM). The composition and kinematics of the ejecta
are sensitive to $M_{\rm He}$.  The lower mass
He-cores with $M_{\rm He} \lesssim 55 ~M_{\odot}$ eject only
the He-envelope. More massive cores eject
a part of the CO layer. The most massive core studied
of $M_{\rm He} = 62 M_{\odot}$ ejects even 
the Si-layer. Such heavy elements may largely
alter the opacity of the CSM.

\noindent (5) 
We examined the 
connections of PPISN, especially the ones with 
massive mass ejection, with the recently observed
Type I super-luminous supernova (SLSN-I) PTF12dam. 
We show that the PPISN model 
produces massive enough CSM, which 
may be able to explain some super-luminous supernovae (including PTF12dam), 
based on the CSM interaction.
The amount of C and O is consistent with the light curve models of SLSNe-I.

\noindent (6)
We compare the masses of 
black holes detected from the gravitational
wave (GW) signals with the black hole masses after the mass ejection of PPISNe. 
Our PPI models predict that the expected black hole masses are
$\sim 38 - 52~M_{\odot}$, i.e., 
the upper limit of the black hole mass is $52~M_{\odot}$.
This is consistent with the current observations.
Some of the events, 
especially GW 170729 which shows a progenitor mass
of $\sim 50 ~M_{\odot}$, could be a remnant left behind by
PPISN. The upper limit of the black hole mass can form
the lower mass limit of the mass gap of the massive black holes
(i.e. the transition from black hole to no-remnant).
Future observations of the black hole mass spectrum 
derived from the merger events of binary
black holes can provide the corresponding constraints 
on such mass limit. The detection of black hole mass prior to
the merger event between $\sim 50$ and $\sim 150 ~M_{\odot}$
can challenge the current black hole formation mechanism 
and its progenitor evolution, and provide insight to the 
implied merging event rate of massive black holes 
evolved from PPISNe.

\noindent (7)
In the future work, we will focus on the observables
of the PPISN in terms of neutrinos and light curves. 
Using our hydrodynamics model,
the expected neutrino signals detected by
terrestrial and the expected light curve will
be calculated. The results will provide a more
fundamental understanding to the properties 
of PPISN, which may be constrained from 
the observables of one the the PPISN candidates.

\noindent (8) In the appendix we show that our results are qualitatively
consistent with the results in the literature,
although some minor differences can be found.

\section{Acknowledgment}

This work has been supported by the World Premier International
Research Center Initiative (WPI Initiative), MEXT, Japan, and JSPS
KAKENHI Grant Number JP17K05382.
S.B. work on PPISN is supported by the Russian Science
Foundation Grant 19-12-00229.
We thank the developers of the stellar evolution code MESA
for making the code open-source.  
We also thank Raphael Hirschi for the insightful discussion
in the stellar evolution of PPISN and his critical comments. 
We at last thank Ming-Chung Chu for his assistance in editing 
the manuscript. 

\software{MESA (v8118; \citep{Paxton2011, Paxton2013, Paxton2015,Paxton2017})}


\appendix

\section{Comparison with Models in the Literature}
\label{sec:comparison}

\subsection{\cite{Yoshida2016}}

In this section we compare our results with some representative
PPISN models in the literature. 

In \cite{Yoshida2016}, the PPISN model of mass
from about 54 $M_{\odot}$ to 60 $M_{\odot}$
(corresponding to a progenitor mass from 
140 $M_{\odot}$ to 250 $M_{\odot}$ in zero
metallicity) are computed. In that work, the calculation 
is separated into two parts. During the 
quiescent and pre-pulsation phases, the 
hydrostatic stellar evolution code is used.
During the pulsation phase, the star model 
is transferred to the dynamical code PPM, 
which follows the expansion of the star
until the mass ejection has ended ($\sim 10^4$ s). 
Then they map the results to the stellar evolution 
code again until the next pulsation.

Their 140 $M_{\odot}$
and 250 $M_{\odot}$ models have similar configurations
as our Models He55A and He60A. 
First, in their 
140 $M_{\odot}$ model (250 $M_{\odot}$ model), 
they observe a total of six (three) pulses
which ejected 3.99 (7.87) $M_{\odot}$ of matter 
before collapse. Model He55A (He60A) 
exhibits three (two) pulses before collapse, 
which ejects 6.78 (8.52) $M_{\odot}$ of matter.
Our models show a smaller 
number of pulses, but give similar ejecta
mass. This means our models can capture the 
energetic pulse well, but not the smaller
pulses.
 
Then we compare the ejection timescales. 
The 140 $M_{\odot}$ (250 $M_{\odot}$) model 
show all pulses within a period of 
0.92 (1434) years, while Model He55A (He60A) 
shows all pulses within a period of 
1341 (2806) years. There is a huge
difference in the pulsation period
in our Model He55A and their 140 $M_{\odot}$ model. 
We notice that the difference comes from the 
strengths of the pulses. In particular, our second
pulse leads to a transition about 100 years
while ejecting $1.45 ~M_{\odot}$. The most similar
event in their model is the fourth pulse, but 
with a transition of only 0.279 year. 

At last we compare the final core composition.
The 140 $M_{\odot}$ (250 $M_{\odot}$) model has
an Fe (CO) core mass at 2.57 (43.51) 
$M_{\odot}$, while in our model, we have
2.49 (38.60) $M_{\odot}$ for the Fe (CO) 
mass. This shows that, despite the difference
in the mass ejection history, our models can 
still capture the major mass ejection events, 
which results in a similar mass ejection
and core composition. However, there is a strong pulse in 
our He55A model, which is not seen in 
their 140 $M_{\odot}$ model. 

\subsection{\cite{Woosley2017,Woosley2019}}

Next, we compare our models with the models from 
\cite{Woosley2017}. We have chosen the PPISN close
to that work; in particular, ours Models
He40A, He50A, He60A and He62A can be 
compared directly with the He40, He50, He60 
and He62 models. In \cite{Woosley2017}, the Kepler code,
which consists of both hydrostatic and hydrodynamics
components, is used to follow the whole
evolution of PPISN. 

First we compare the mass ejection history. 
In \cite{Woosley2017}, there are
9, 6, 3 and 7 pulses
with a total mass loss of 0.97, 6.31, 12.02 and
27.82 $M_{\odot}$ for Models He40, He50, 
He60 and He62 respectively.
In our models, we have 6, 3, 2 and 2 pulses with a
total mass loss of 2.22, 2.61, 9.52 and 12.85
$M_{\odot}$ for Models He40A, He50A, 
He60A and He62A respectively.
Again, our code tends to produce fewer 
pulses and the pulses in general eject
fewer matter. One of the differences is
how shock is treated. For a shock-capturing scheme
with a larger dissipation, the kinetic energy will
be partly dissipated into thermal energy, such that 
the star is globally thermalized instead of ejecting
matter through kinetic pulses. 
Another origin of the differences can be 
related to the nature of the instability
of PPISN. Since the trigger of the explosive O-burning 
comes from the pair-instability, which is 
very sensitive to the initial condition (e.g.
how we evolve the stellar evolution model 
before the pulsation and between pulses)
and numerical treatment (e.g. how convection 
and mass ejection are treated). For example,
a stronger contraction can lead to more O-burning
in the core, which gives much stronger pulsation
and hence more mass loss. In fact, such 
dependence can also be seen in other 
field. For example, in the propagation of flame, 
since it is unstable towards hydrodynamics instability, 
\citep{Glazyrin2014}. The burning history can be 
highly irregular in the unstable regime. 
 
Next we compare the timescale of the pulsation.
In this work, the whole pulsation until 
collapse last for 0.38, 61.3, 2806 and 
6610 years for the four models, while in \cite{Woosley2017}
they are $2.48 \times 10^{-3}$, 0.38, 2695
and 6976 years. It shows that for massive He cores, 
our results agree with their work but there are
large differences when the He core becomes
less massive. In that case, our final pulse is 
always strong enough to re-expand the star 
again before the final collapse, which significantly
lengthens the pulsation period. 

Then we compare the Fe core mass. In \cite{Woosley2017}
the core has 2.92, 2.76, 1.85 and 3.19 $M_{\odot}$
Fe. In our models, we have 3.42, 1.73, 1.64 and 2.66 $M_{\odot}$.
There is a dropping trend from Model He40A to 
He60A, which corresponds to the trend that the 
pair-instability occurs at a lower density when the 
mass increases. On the other hand, near
the pair-instability regime, the pulsation becomes
sufficiently vigorous which enhances the NSE-burning.

At last we compare the explosion energy. We compare the 
Model He62A, which has the largest explosion energy. 
In our model, in the second big pulse, the star has 
its total energy increased by $2.0 \times 10^{51}$ erg
while the maximum kinetic energy achieved is 
$2.8 \times 10^{51}$ erg. This is very similar to 
the result in \cite{Woosley2017}, where
the pulse is observed to have a kinetic energy
of $2.8 \times 10^{51}$ erg.  

One major difference we notice is in the pair-instability
limit, for Model He64A, our model shows a 
higher explosion energy. Across the strongest pulse, 
there is a change of total energy by $1.6 \times 10^{52}$ erg,
where the maximum kinetic energy of the system
is $\sim 1.7 \times 10^{52}$ erg. In \cite{Woosley2017} 
the kinetic energy is reported to be $4 \times 10^{51}$ erg. 
We observe that the difference comes from the 
number of pulsation, where our Model He64A has two 
big pulses but only one in their work. The first pulse
has incinerated the $^{16}$O in the core while ejecting 
on the surface. This means that the star has to 
reach a more compact state before the star can 
explode. As a result, the amount of energy produced
in the exploding pulse is much larger. 
 
Our results
show a systematically lower number of pulses
with slightly lower ejecta mass. The pulsation periods
qualitatively agree with each other except for 
models with a final strong pulse, which may
significantly lengthen our pulsation period. 
Also, in our explosion models, the system tends
to store the energy in terms of internal energy
instead of kinetic energy, as a result, 
the star tends to expand globally, where 
the excess energy and momentum of the star
is transferred mostly to the surface. This 
ejects the low density matter and leaves
a bounded and hot massive remnant. 
Despite the differences in the pulsation,
globally the nucleosynthesis agrees with each 
other because most of the heavy elements are produced
by the strong pulses, where our results are 
consistent with those in the literature. 

In \cite{Woosley2019}, the He star models are further evolved
with mass loss. The solar metallicity of the Fe group is assumed in the mass loss rate for He stars.
We note that this is not consistent with the mass loss history from
the main-sequence, because the He core of the solar metallicity star
becomes too small to undergo PPI (see our Figure \ref{fig:HeCore_plot}).
The final black hole mass is lower
compared to his previous work. We notice that 
the PPISN models in this work have in general stronger 
pulsations. A 50 (60) $M_{\odot}$ He-core ejects 
$\sim 7 $(57) $M_{\odot}$, which is much higher than 
$\sim 6$ (12) $M_{\odot}$ mass ejection in his previous work.
Our results are closer to his previous work.
This might depend on the thickness of the He layer
which is determined by the mass loss history.

\subsection{\cite{Marchant2018}}

This work is one of the recent work which uses the same MESA
code (Version 11123) \citep{Paxton2017} to evolve 
the evolutionary path of PPISN. Their work has a similar
setting to this work. Here we briefly compare their results
with our results.

They have computed an array of single star models
from 40 -- 240 $M_{\odot}$ with semi-convection, 
Riemann solver using the HLLC solver
and the $approx21$ nuclear reaction network. 
They treat the mass loss of the star by considering
the average escape velocity. 

Our work agrees qualitatively with theirs. For
a lower mass He core model, some distinctive differences can be seen. 
For example, in their models, 
multiple pulsations are observed. They observe
a total of 4 pulses for the 54 $M_{\odot}$ stars (corresponds
to 39.73 $M_{\odot}$ at He depletion). On the other hand, our 40 $M_{\odot}$
He star model gives a total of 6 pulses. They observe in total 0.63 
$M_{\odot}$ mass ejection before collapse while ours is about 
2.2 $M_{\odot}$. The duration in their model is shorter ($\sim 7 \times 10^{-4}$ year)
but ours is longer ($\sim 0.02$ year) after the onset of pulsation. 
For a higher mass He core model such as 
$\sim 87 ~M_{\odot}$ (corresponds to 60.04 $M_{\odot}$ 
at He depletion). They show only 2 pulses, which is the 
same as our 60 $M_{\odot}$ He star model. The duration
also agrees with each other (their model shows a duration of 
$7.5 \times 10^3$ years while ours is shorter at $3 \times 10^3$ years. 
A total of 4.6 $M_{\odot}$ mass loss is found in their model (the 
pre-He depletion mass loss is excluded) while ours is at a higher value
$\sim 8.5 ~M_{\odot}$.  

We notice that their models and our models do not completely
agree with each other. 
We notice that there are some critical differences in the implementation
of this work from their work. First, they consider the evolution
of the H-free stars, with a metallicity at 0.1 $Z_{\odot}$.
The He-core mass is therefore a function of the progenitor
mass, instead of a direct model parameter as controlled
in our models. 
Furthermore, 
they use the Riemann solver (HLLC) in the newer version
instead of the artificial viscosity scheme. How the 
pulse transfers into shock at the near-surface area can
be different.

\section{Effects of Hydrostatic Convective Mixing}
\label{sec:conv_mix}

\begin{figure}
\centering
\includegraphics*[width=7cm,height=5cm]{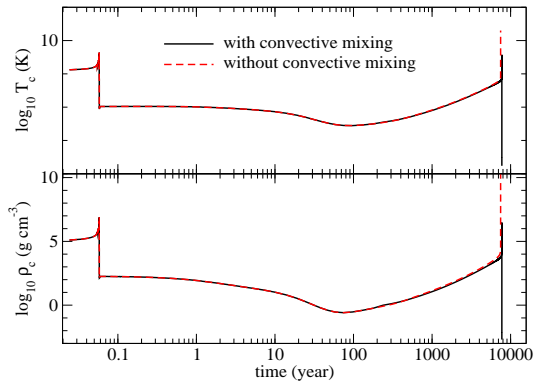}
\caption{(upper panel) The central temperature against time 
for Model He60A with and without convection. (lower panel)
Similar to the upper panel, but for the central densities.}
\label{fig:tempc_rhoc_time_He60_conv_plot}
\end{figure}

\begin{figure}
\centering
\includegraphics*[width=7cm,height=5cm]{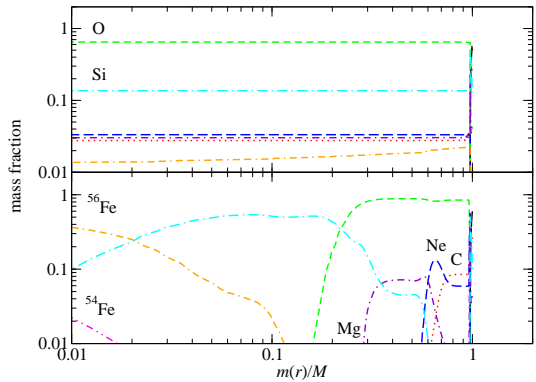}
\caption{The chemical abundance profiles for Model He60A prior
to its second contraction at a central temperature 
$\approx 10^9$ K with convection (upper panel) and 
without convection (lower panel).}
\label{fig:xiso_He60_conv_plot}
\end{figure}

In \cite{Woosley2017} the PPISN is prepared for models
with convective mixing. It is mentioned that the convective 
mixing is essential to evolve the star correctly to readjust the 
chemical composition of the remnant. It is unclear how much 
the convective mixing can change the evolutionary path of the
PPISN. Here we compare the model of He60A by treating 
the convective mixing as an adjustable parameter. 
In Figure \ref{fig:tempc_rhoc_time_He60_conv_plot} we plot the
central temperature (upper panel) and central density (lower panel)
against time for Model He60A for both choices. It can be seen that
the effects of convective mixing are huge. In the model with mixing 
switched on, in the second pulse it leads to a large amplitude 
expansion, which leads to significant mass loss afterwards 
before its third contraction to its collapse. On the 
other hand, the model without convective mixing has a faster
growth of central temperature and central density, where
the star collapses without any pulsation.

To understand the difference, we plot in Figure \ref{fig:xiso_He60_conv_plot}
the chemical composition of the star before the second
contraction takes place. We pick both star models when it 
has a central temperature of $10^9$ K. It can be seen that
the role of convective mixing is clear that the mixing
not only re-distribute the energy of the matter, the composition
in the large-scale is modified. A considerable amount of fuel
is re-inserted into the core, which contains O and Si from the 
unburnt envelope, and some remained $^{54}$Fe and $^{56}$Fe produced
in the first contraction. This shows that the convection during the expansion 
is important for the future nuclear burning to correctly predict
the strength of the pulse, which affects the nucleosynthesis as
well as the mass loss. 

\begin{figure}
\centering
\includegraphics*[width=7cm,height=5cm]{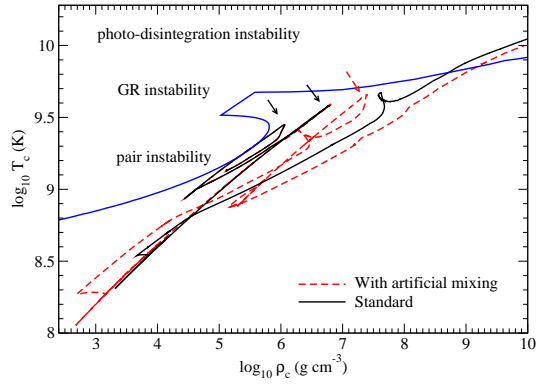}
\caption{The central temperature against central density in log$_{10}$
scale for two models, the He60A with standard mixing scheme
(See also Section \ref{sec:methods}) and with artificial 
enforced mixing scheme. The blue lines and the arrows follow the same meaning as in Figure \ref{fig:tempc_rhoc_plot}. The arrows
(black solid arrows and red dashed arrow) are the moments
of the pulsation for the tested models.}
\label{fig:Tc_rhoc_artmix_plot}
\end{figure}

To further demonstrate the importance of convective mixing 
to the strength and number of pulsations, we perform some
contrasting study of two models, one is Model He60A
the other is similar to He60A, but with enhanced mixing. 
We have shown in Section \ref{sec:hydro} from the Kipperhahn
diagram that the convection mixing in Model He60A is less strong
that during its quiescent phase after pulsation, the star does
not exhibit the global convective phase, unlike other models
like He40A, He50A and He62A. So, this model becomes a good candidate
to demonstrate the effects of convective mixing between pulsation.
To provide the enhanced mixing, we enforce the whole star to
undergo mixing process during its expansion and when it is 
fully relaxed. We defined the critical temperature be $10^9$ K
below that the star is fully relaxed for convective mixing. 

In Figure \ref{fig:Tc_rhoc_artmix_plot} we plot the central
temperature against central density (both in logarithmic scale)
for the two models. The evolution of He60A is exactly the same as
that presented in previous section. Here, we look into more details
for the model with artificial mixing. Before the second pulse, the two 
models exhibit exact the same trajectories. It is because the central
temperature has barely reached below $10^9$ to trigger the mixing.
But after the second pulse, which has mass ejection, its 
central temperature goes below $10^9$ K. The one with 
enhanced mixing, because it involves mixing material with 
the outer elements, which has in genearl lower temperature and 
lower atomic mass, it can reach a low central temperature
during its expansion. Also, the mixing process brings in 
the C- and O-rich material into the core. In the third contraction,
unlike the "standard" model presented in the main text, the core 
exhibits the third pulsation. However, the strength is not 
strong enough to trigger mass loss on the surface. Then, 
although the core reaches one more time below the $10^9$ for
the hand-made convective mixing, the O-abundance of the star
becomes too low that the core becomes massive enough to collapse
directly, without triggering the fourth pulsation. 

\section{Effects of Artificial Viscosity}
\label{sec:art_vis}

\begin{figure}
\centering
\includegraphics*[width=7cm,height=5cm]{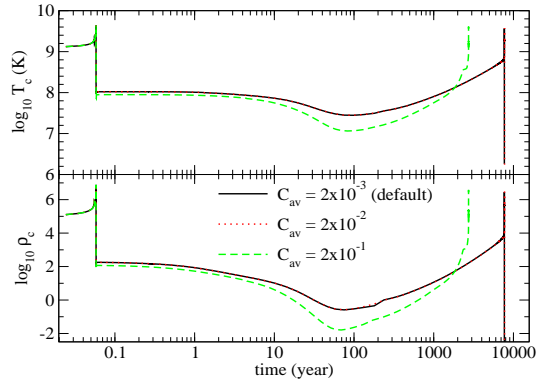}
\caption{(upper panel) The central temperature against time 
for Model He60A with different levels of artificial 
viscosity. (lower panel)
Similar to the upper panel, but for the central densities.}
\label{fig:tempc_rhoc_time_He60_av_plot}
\end{figure}

Another important parameter in numerical hydrodynamics modeling 
is the artificial viscosity. Owing to the lack of Riemann solver
(exact or approximate) for the spatial derivative, 
artificial increases of pressure is needed to prevent
the shock from over-clumping the mass shells. However, the artificial
viscosity formula contains one free parameters $C_{{\rm av}}$. 
The default value from the package 'ccsn' in the \software{MESA} test suite 
is $C_{{\rm av}} = 2 \times 10^{-2}$. To probe the effects of
this parameter, we carry out a control test by varying 
$C_{{\rm av}}$.

In Figure \ref{fig:tempc_rhoc_time_He60_av_plot} the time dependence of 
the central temperature
(upper panel) and central density (lower panel)
are plotted for Model He60A with 
$C_{{\rm av}} = 2 \times 10^{-3}$, $2 \times 10^{-2}$ (default value)
and $2 \times 10^{-1}$. Results with
$C_{{\rm av}} = 2 \times 10^{-3}$ and $2 \times 10^{-2}$ 
are almost identical. This shows that the default choice
of $C_{{\rm av}}$ can maintain the 
shock propagation and produce convergent results. 
On the other hand, when $C_{{\rm av}} = 2 \times 10^{-1}$, very different
outcome appears. The first expansion has reached to a lower 
central temperature and density. Furthermore, the two quantities
are in general lower than the cases with lower $C_{{\rm av}}$
during the expansion. The second contraction also takes place
a few thousand years before the other two cases. This shows that 
if a too large artificial viscosity is chosen, the pressure heating
also alters the shock heating and its associated nuclear burning
in the star, thus affecting the consequent configurations. 

\section{Effects of Hydrodynamics Convective Mixing}
\label{sec:dyn_mix}

In the main text we have mentioned that the role of convective mixing 
is less important in the hydrodynamics during shock outbreak 
because the typical timescale 
of convective mixing is longer than the hydrodynamical timescale. 
However, it remains interesting to examine when convective 
mixing is included, how it changes the evolutionary path. 
In fact, a more consistent and accurate approach to follow the 
evolution requires the input of convective mixing, but it always 
induces numerical instabilities which impedes any further evolution.
Here we attempt to study how convective mixing affects the pulsation
history of PPISN.

\begin{figure}
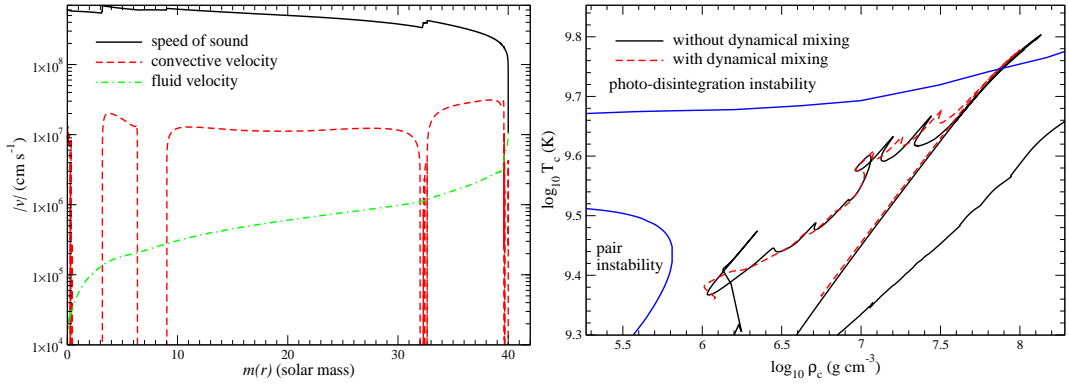

\centering
\includegraphics*[width=7cm,height=5cm]{fig39a.eps}
\includegraphics*[width=7cm,height=5cm]{fig39b.eps}
\caption{(left panel) The fluid velocity, speed of sound and 
convective velocity of Model He40A at $T_c = 10^{9.6}$ K. 
(right panel) The thermodynamics trajectory of Model He40A with 
dynamical mixing (red dashed line) and without dynamical mixing (black solid
line). The blue lines and the arrows follow the same meaning as in 
Figure \ref{fig:tempc_rhoc_plot}.}
\label{fig:conv_plot}
\end{figure}

In the left panel of Figure \ref{fig:conv_plot} we plot the speed of sound, 
fluid velocity and convective velocity for the Model He40A
when it is rapidly contraction before the first pulse at
a central temperature of $10^{9.6}$ K. We can see that, indeed,
the convective velocity is about $\sim 1 \%$ of the speed of sound,
while the fluid velocity is less than $10^{-4}$ of the speed of 
sound. The star is close to hydrostatic equilibrium, in contrast
to the massive star $\sim 60 M_{\odot}$ counterpart. The more
compact structure of the star also means a shorter convection
timescale. So, mixing can be influencing to the pulsation process.

As mentioned switching on convection can be problematic in the hydrodynamics.
To bypass this difficulties, instead of doing mixing in the 
hydrodynamics, we post-process at every step the abundance
profile to mimic the mixing process. Similar to the standard
mixing length theory procedure, we first locate the mass shells
which can undergo convection. Then we calculate the convection
velocity and the corresponding mixing timescale $t_{{\rm mix}}$. After that,
we compare with the timestep $\Delta t$. If $t_{\rm mix} < \Delta t$,
complete mixing is assumed; otherwise partial mixing among the 
cells in the convection zone is assumed. We notice that a consistent way to 
do the mixing process requires mixing entropy too. However, 
this affects the pressure which in terms affects the dynamics. 
In fact, it is the mixing of fuel to the actively burning site
important for the trigger of pulsation. As a first approximation,
we neglect this complication.

In the right panel of Figure \ref{fig:conv_plot} we plot the 
thermodynamics trajectories of Model He40A using the default prescription
(no dynamical mixing) and the described mixing process. 
We can see that both curves are very similar qualitatively. 
However, minor changes can be seen by the small scale pulses
in the star. Model with dynamical mixing has fewer small pulse.
The moment where the large pulse takes place 
differs. The model without dynamical mixing occurs at a higher 
$T_c \sim 10^{9.8}$ K while that with dynamical mixing occurs
at a lower $T_c \sim 10^{9.75}$ K. One possibility for this difference
is that for small pulse, the mixing tends to lower the $^{16}$O 
abundance available to the active burning site, which is more
local. On the other hand, the mixing allows more zone to be 
rich in $^{16}$O when the star needs to carry out a collective
burning of $^{16}$O. Therefore, it can occur earlier.
Depsite the difference it shows that the mixing process is 
efficient to the lower mass PPISN but the replenishment of 
fuel does not particularly enhance the pulsation process.

\newpage

\bibliographystyle{apj}
\pagestyle{plain}
\bibliography{biblio}

\end{document}